# Modeling the Ferroelectric Phase Transition in Perovskite Relaxors and New Corresponding Solving Method

Yi-Neng Huang[1,2*] and Li-Li Zhang[1,2]

1. Xinjiang Laboratory of Phase Transitions and Microstructures in Condensed Matters, College of Physical Science and Technology, Yili Normal University, China

2. National Laboratory of Solid State Microstructures, school of Physics, Nanjing University, China

**Abstract** Three-dimensional-random-site-Ising-model (3D-RSIM) along with the Glauber dynamics of pseudo-spins (PSs) is applied for the first time to study the relaxion-ferroelectric-phase-transition (RFPT) of perovskite-relaxor-ferroelectrics (PRFEs) in detail. To solve this model, we proposed a new method, mean-field of PS-strings (MF-PSSs), which includes the effect of the interfaces between the groups with or without PSs. We find: 1) 3D-RSIM is a mixed system consisting of permanent-paraelectric-subsystem (PPSS), low-transition-temperature ferroelectric-subsystem (LTT-FSS), and high-transition-temperature ferroelectric-subsystem (HTT-FSS). The contents of these three subsystems change with the PS concentration ($\phi$) of the model; 2) When $\phi < \phi_p$ ($\phi_p$ is the percolation threshold of the 3D-RSIM), HTT-FSS dominates the whole system. On cooling, the system undergoes an inhomogeneous-diffuse-ferroelectric-phase-transition (IDFPT), and as $\phi$ increases, it shows a crossover from the critical, gradually to Vogel-Fulcher, and to Arrehnius type slowdown. Specifically, we observe this crossover from the relationship between the peak temperature ($T_m$) of the real susceptibility of the correlated relaxation of PSs and angular frequency ($\omega$); 3) When $\phi = \phi_p$, the three subsystems have comparable contents. However, when $\phi > \phi_p$, PPSS becomes the dominating subsystem. As temperature decreases, the spontaneous polarization and specific heat of the whole system show an IDFPT, but $T_m$ shifts to lower temperature with $\omega$ until 0K. Since the relaxation time of PPSS rapidly increases at lower temperature and diverges at 0K, during a cooling process at a limited rate, PPSS must be frozen at a certain temperature, i.e. PS glass transition occurs. In short, with the increase of $\phi$, 3D-RSIGM gradually evolves from the IDFPT system to PS glass system, and $\phi_p$ is the characteristic concentration of this evolution. Moreover, based on the coupling between PSs and crystal lattice, we also give a new possible mechanism of Burns transformation. The model predictions in this paper are consistent with the corresponding Monte Carlo simulations and typical PRFE experimental results.

---

[*] Email: ynhuang@nju.edu.cn

## 1. Introduction

Since the discovery of relaxor-ferroelectrics (RFEs) by Smolenskii and co-workers in 1954 [1-3], RFEs have been often studied by many researchers [4-16] due to their (potential) important applications based on their huge susceptibility, piezoelectric coefficient... [5-7] However, unlike the conventional-ferroelectric-phase-transition (CFPT), the mechanism of relaxion-ferroelectric-phase-transition (RFPT) is not well understood [4-25].

Among the existing RFPT theories, there are some main influencing ones. In 1970, Smolenskii calculated the distribution of microscopic component concentration, proposing a composition disorder model to calculate the static susceptibility of RFEs based on an assumption that the local phase transition temperature is proportional to the microscopic concentration [3]. Random field theory studied by Westphal et al. emphasized the influence of such field to the phase transition and domain structures, but this theory is lack of quantitative predictions to compare with experiments [17-20]. In the early 1990s, Viehland et al. suggested a polar glass model based on the coupling between polar-nanoregions (PNRs), but they did not give a clear model Hamiltonian and its solutions [21-23]. The spherical random-field random-bond theory by Blinc et al. treated PNRs as effective dipoles, assuming they are in a random Gaussian internal field with their interactions also following a Gaussian distribution. This theory successfully predicted the ferroelectric phase transition and effective dipole glass transition with specific parameters. However, it has an unphysical assumption that the effective dipoles have an infinite long and strong interactions [24,25]. To make this theory better, we can further calculate the dynamics especially the complex susceptibility of the system, as well as writing down the distributions of the PNRs, the random field, and random bond on a microscopic level.

RFEs is mainly discovered in perovskite-relaxor-ferroelectrics (PRFEs). In section 2, we first apply the three-dimensional-random-site-Ising-model (3D-RSIM) along with the Glauber dynamics of pseudo-spins (PSs), referred as 3D-RSIGM, to study the RFPT in PRFEs in detail. In section 3, a new method of solving 3D-RSIGM is proposed, which is the mean-field of PS-strings (MF-PSSs) accounting for the effect of the interfaces between the groups with or without PSs. In section 4, we show computer simulations to calculate the distributions of the length of PSSs and their interaction with their nearest neighbors, and then we further calculate the order parameter, specific heat and complex susceptibility of the PSSs and 3D-RSIGM. Moreover, based on the coupling between PSs and crystal lattice, we also give a new possible mechanism of Burns transformation. In section 5, we compare our model predictions to related Monte Carlo simulations and experimental results, and they show good agreements. Later in the same section, we also propose some potential future study of this model to make it better. In the last section, we give the conclusion of this paper.

## 2. Simplified Model of the RFPT in PRFEs

PRFEs are systems in which two or more ions are unevenly distributed on the crystal lattices, also known as compositional disorder systems, for example, $Sc^{3+}$, $Ta^{5+}$ in $PbSc_{1/2}Ta_{1/2}O_3$; $Mg^{2+}$, $Nb^{5+}$ in $PbMg_{1/3}Nb_{2/3}O_3$ ($PM_{1/3}N_{2/3}$); $Zr^{4+}$, $Ti^{4+}$ in $BaZr_xTi_{1-x}O_3$ ($BZ_xT_{1-x}$) [26-28].

According to the disorder ion valence, it is divided into homovalent (such as $BZ_xT_{1-x}$) and heterovalent (such as $PM_{1/3}N_{2/3}$ and $Sr_xBa_{1-x}Nb_2O_6$) PRFEs. Both random-internal-stress-field (RISF) and random-internal-electric-field (RIEF) are generated due to the differences in the size and charge of the disorder ions in the heterovalent PRFEs. However, there is only RISF in the homovalent ones [18].

Taking into account the fact that the heterovalent PRFEs (such as $PM_{1/3}N_{2/3}$ and $Sr_xBa_{1-x}Nb_2O_6$) and the homovalent ones ($BZ_xT_{1-x}$) have qualitatively the same characteristics of complex susceptibility [29-39], we could conclude that RIEF has no qualitative influence on the characteristics of RFPT (Of course, there must be some quantitative effects). In other words, RIEF can be ignored in a simplified microscopic model of RFPT. Considering the similarity between RIEF and RISF, RISF can also be neglected in the simplified model.

PRFEs such as $BZ_xT_{1-x}$ [37] and $Sr_xBa_{1-x}Nb_2O_6$ [35] show that with the increase of x, the system evolves from the occurrence of CFPT to RFPT on cooling. Therefore, it is feasible to obtain a reasonable microscopic model of RFPT by appropriately introducing component disorder into the existing models of CFPT.

There are two successful schemes to describe CFPT, including pseudo-spin (PS) (Equivalence of the orientation movement of permanent dipoles to spin) [40], and soft-mode [41-44]. Here, we intuitively choose the first one because it would be convenient and direct to introduce the composition disorder. The PS scheme includes three-dimensional Ising-model (3D-IM) and its extended ones. The following takes $BZ_xT_{1-x}$ as an example [45-49] to construct 3D-IM with disorder components, i.e. 3D-random-site-IM (3D-RSIM) [50, 51] of PRFEs.

According to the experimental results, the paraelectric-ferroelectric phase transition temperature ($T_{c1}$) of $BaTiO_3$ (BT) is approximately 380K, the Curie-Weiss constant of the paraelectric phase ($C_w$) $\approx 1.6 \times 10^5$K [37]. There is not any phase transition of $BaZrO_3$ (BZ) from 2K to 1375K, and the Curie constant of BZ ($C_c$) $\approx 3.2 \times 10^3$K [47]. According to the Weiss mean field of 3D-IM, the ferroelectric interaction energy ($J$) between PSs can be obtained, and $J(BT) \approx \frac{T_{c1}}{6} = 63$K, $J(BZ) \approx 0$K; the ratio of the permanent electric dipole moments of $\mu(BZ)$ to $\mu(BT)$ is: $\frac{\mu(BZ)}{\mu(BT)} = \sqrt{\frac{C_c}{C_w}} = 0.14$, so as a primary approximation, we can assume $\mu(BZ)$ to be zero in a microscopic model. Therefore, $BZ_xT_{1-x}$ can be simplified as the following 3D-RSIM [50,51]:

1) The crystal lattice structure of the model is simple cubic, and the permanent electric dipoles (permanent dipole moment being $\mu$) of concentration (1-$\phi$) randomly distribute on the lattice points. The rest of the lattice points are empty.

2) The orientation motion of the permanent electric dipoles is equivalent to two states of PSs. In this paper, $\sigma_i$ is used to represent the i$^{th}$ PS in the model, and its two states are represented by $\sigma_i = \pm 1$ ($i = 1, 2 \cdots N(1 - \phi)$, where $N$ is the total number of the lattice points with $N \rightarrow \infty$). The lattice points without permanent dipoles are equivalent to having no PS.

3) Only the interaction between the nearest-neighbor PSs is not zero, and the interaction energy between the i$^{th}$ and its nearest-neighbor j$^{th}$ PS is $-J\sigma_i\sigma_j$.

4) All PSs in the model are in a heat bath consisting of ions on the crystal lattice at temperature $T$.

Therefore, the Hamiltonian of the model is,

$$H_{RSIM} = -J \sum_{i \neq j}^{\{nn\}} \sigma_i \sigma_j r_i^\phi r_j^\phi \tag{2.1A}$$

where $r_i^\phi$ is a random function that $r_i^\phi = 0$ for $r < \phi$, $r_i^\phi = 1$ when $r \geq \phi$, and $r$ is a randomly generated number between 0 and 1. $\{nn\}$ represents the summation of all the nearest-neighbors.

Moreover, in order to describe the dynamic parameters such as complex susceptibility of the model, we use the Glauber transition probability ($w(\sigma_i)$) [52-54] from $\sigma_i$ to $-\sigma_i$ in unit time (appendix 1) is,

$$w(\sigma_i) = \frac{\nu}{2}\left[1 - \sigma_i \tanh\left(\frac{H_i}{k_B T}\right)\right] \tag{2.1B}$$

where $H_i \equiv J \sum_j^{\{nni\}} \sigma_j r_j^\phi$ is the local field of $\sigma_i$, $\{nni\}$ represents the summation of the nearest-neighbors of $\sigma_i$; $\nu = \nu_0 \exp\left(-\frac{U_B}{k_B T}\right)$, $U_B$ is the energy barrier that PSs stride over during the transition from $\sigma_i$ to $-\sigma_i$, and $\nu_0$ is the orientation vibration frequency of PSs in their local energy valleys. For convenience, 3D-RSIM with the Glauber transition probability is called as 3D-random-site-Ising-Glauber-model (3D-RSIGM).

The reasons for choosing Glauber transition probability are: 1) The Weiss mean-field form of 3D-RSIGM with $\phi = 0$ is same as Mason theory [55] describing the critical relaxation of 2$^{nd}$ order CFPT [56], which is nearly consistent with experiments; 2) For the correlated relaxation of multi-PSs, the MF-PSSs of 3D-RSIGM when $\phi \to 0$ gives the similar critical slowdown of relaxation time to the theory (section 4.3 of this article).

We can imagine that for BZ$_x$T$_{1-x}$, $\phi \approx x$, $J$, $\mu$, and $U_B$ are almost irrelevant to x when x is small. However, when x is large enough, $\phi >$ x, $J$, $\mu$, and $U_B$ will decrease as x increases due to the RISF. In addition, $PM_{1/3}N_{2/3}$, $PbZn_{1/3}Nb_{2/3}O_3$ and $PbMg_{1/3}Ta_{2/3}O_3$ correspond to 3D-RSIGM of $\phi \approx 1/3$. For the more complicated case of $Sr_xBa_{1-x}Nb_2O_6$, the relationship between $\phi$ and x still needs to be studied.

It is worth noting that 3D-RSIM is recognized as one of the realistic models of spin glasses (such as $BiSc_xMn_{1-x}O_3$) [57-62]. Taking into account the similarity of physical quantities of PRFEs and spin glasses with temperature and frequency, it can be considered that 3D-RSIM is also one of the realistic simplified models of PRFEs.

The solution of 3D-RSIGM consists of calculating: 1) the thermodynamic quantities of 3D-RSIM, such as order parameters, static susceptibility, and specific heat; 2) the dynamic parameters of 3D-RSIGM, such as complex susceptibility.

## 3. New Mean-Field of PS-Strings to Solve 3D-RSIGM

According to the author's knowledge, the exact partition function of 3D-RSIM (including 3D-IM) have not yet been obtained. The approximate solving methods are renormalization group theory (RGT) [63,64] etc. Below, we illustrate the difficulties to solve 3D-RSIM from simple examples.

It can be imagined that in 3D-RSIM for $N \to \infty$, the following two subsystems must be included:

1) Finite one-dimensional-Ising-model (1D-IM) [65] of PSs. The exact solution shows that 1D-IM is always at paraelectric state [66-67]. This article refers to this kind of subsystem as the permanent-paraelectric-subsystem (PPSS).
2) Finite two-dimensional-Ising-model (2D-IM) of PSs. Its exact solution indicates that a diffuse peak of specific heat appears at a certain temperature [68, 69], and the total spontaneous polarization is zero [70], which leads to the conclusion that there is not any phase transition in finite 2D-IM. However, according to Imry-Ma theory [71], this originates from the boundary effect of finite systems, which leads to the multi-domain formation in ferroelectric phase. Now, it is widely believed that there is a diffuse-ferroelectric-phase-transition (DFPT) in finite 2D-IM. Here, this kind of subsystem is called as ferroelectric-subsystem (FSS).

3D-RSIM includes both PPSSs and FSSs, and the phase transition temperature of FSS has a distribution. Therefore, in addition to the physical quantities, such as order parameter, specific heat, and susceptibility of CFPT in homogeneous systems, the description of 3D-RSIM also requires the contents of different subsystems.

The coexistence of PPSS and FSS in 3D-RSIM also indicates that RGT [63,64] has the following problems: 1) For $\phi < \phi_p$ ($\phi_p \approx 0.69$, the percolation threshold) [72], the model has a 2$^{nd}$ order CFPT with temperature according to this theory. This obviously ignores PPSS, especially when $\phi$ is close to $\phi_p$; 2) Based on RGT, the model has no phase transition when $\phi \geq \phi_p$, which clearly neglects the phase transition of FSS.

The finite series expansion method for solving 3D-IM is only accurate at either low or high temperatures, and the calculation error is larger near the phase transition. The coexistence of PPSS and FSS in 3D-RSIM will further increase the error of this method [73].

In this paper, we propose a new mean-field of multi-PSs, i.e. mean-field of PS-strings (PSSs) to solve 3D-RSIGM, and it is called MF-PSSs to distinguish from the mean-field of straight spin chains for solving 3D-IM [74]. The definition of a PSS is that any internal PS is connected to two nearest neighbor PSs, and the endpoint PS only to one nearest-neighbor PS in the string. Here, a PSS containing $n$ PSs is expressed as n-PSS. MF-PSSs includes:

1) PSS construction in 3D-RSIM: (1) Connect the nearest-neighbor PSs into short PSSs along the z-axis direction of the crystal lattice; (2) Any two nearest neighbor endpoints of the short PSSs are connected along the y-axis direction (An endpoint already connected to a PSS is no longer reconnected); (3) Continue to connect any two nearest neighbor endpoints along the x-axis direction to form long PSSs. This construction process is schematically shown in Fig.1.

2) Count the number of n-PSSs with the string length ($n$) in the model, obtaining the corresponding distribution function ($q_n$).

3) Count the number of n-g-PSSs (n-PSSs with the nearest neighbor number of PSs being $g$) in the model, getting the corresponding distribution function ($\rho_n^g$).

4) The interaction between n-g-PSSs and their nearest-neighbor PSs is described by the following mean-field of Weiss type [75] (with considering the interfacial effect between the groups with or without PSs [69]), $-J\frac{g}{n}\left[\left(1-\frac{1}{n}\right)\eta_n^g + \frac{1}{n}\right]\sum_{i=1}^{n}\sigma_i$ (see Appendix 2 for details), where $\sigma_i$ indicates the $i^{th}$ PS states of n-g-PSSs, and,

$$\eta_n^g \equiv \frac{1}{n}\sum_{i=1}^{n} s_{ni}^g \tag{3.1}$$

which is the order parameter of n-g-PSSs, and $s_{ni}^g$ is the expectation value of $\sigma_i$ (Appendix 3-4).

Therefore, the Hamiltonian of n-g-PSSs is,

$$H_n^g = -J\sum_{i=1}^{n-1}\sigma_i\sigma_{i+1} - J\frac{g}{n}\left[\left(1-\frac{1}{n}\right)\eta_n^g + \frac{1}{n}\right]\sum_{j=1}^{n}\sigma_j \tag{3.2A}$$

From $H_n^g$, the Glauber transition probability $w(\sigma_i)$ [52-54] of n-g-PSSs is obtained (Appendix 1):

For $n = 1$,

$$w(\sigma_1) = \frac{v}{2}(1 - \gamma\sigma_1) \tag{3.2B}$$

For $n = 2$,

$$\begin{cases} w(\sigma_1) = \frac{v}{2}[1 - \gamma\sigma_1 + \alpha(\gamma - \sigma_1)\sigma_2] \\ w(\sigma_2) = \frac{v}{2}[1 - \gamma\sigma_2 + \alpha(\gamma - \sigma_2)\sigma_1] \end{cases} \tag{3.3C}$$

For $n \geq 3$,

$$\begin{cases} w(\sigma_1) = \dfrac{V}{2}[1 - \gamma\sigma_1 + \alpha(\gamma - \sigma_1)\sigma_2] \\ w(\sigma_n) = \dfrac{V}{2}[1 - \gamma\sigma_n + \alpha(\gamma - \sigma_n)\sigma_{n-1}] \\ w(\sigma_k) = \dfrac{V}{2}[1 - \gamma\sigma_k + \beta(\gamma - \sigma_k)(\sigma_{k-1} + \sigma_{k+1})] \end{cases} \quad (3.3D)$$

where $k = 2, \cdots n - 1$, $\alpha \equiv \tanh(\varpi)$, $\beta \equiv \dfrac{1}{2}\tanh(2\varpi)$, $\varpi \equiv \dfrac{\Theta_J}{T}$, $\Theta_J \equiv \dfrac{J}{k_B}$, $\gamma \equiv \tanh(\theta)$, $\theta \equiv \dfrac{\Theta_n^g}{T}\left[\left(1 - \dfrac{1}{n}\right)\eta_n^g + \dfrac{1}{n}\right]$, $\Theta_n^g \equiv \dfrac{g}{n}\Theta_J$, $k_B$ is Boltzmann constant.

## 4. Order parameter, specific heat, complex susceptibility, Burns transformation of 3D-RSIGM

This section mainly includes: 1) The number of n-PSSs in 3D-RSIM was counted by computer simulation, and the results of $q_n$ and $\rho_n^g$ were obtained in subsection 4.1; 2) In subsection 4.2, according to Eq.3.2A, we first strictly calculate the order parameter ($\eta_{ne}^g$), static susceptibility ($\chi_s^{ng}$) and specific heat ($c_n^g$) of n-g-PSSs, then obtain the order parameter ($\eta$), spontaneous polarization ($P_s$), static susceptibility ($\chi_s^{ps}$), specific heat ($c_{ps}$) of 3D-RSIM by these results, $q_n$, and $\rho_n^g$, and get the contents of PSS and FSS with $\phi$ in the model; 3) In subsection 4.3, the complex susceptibility ($\chi_n^{g*}$) of n-g-PSSs was first strictly calculated according to Eq.3.2, then the complex susceptibility ($\chi_{ps}^*$) of 3D-RSIGM is obtained based on $\chi_n^{g*}$, $q_n$ and $\rho_n^g$; 4) A new possible mechanism of Burns transformation is given in subsection 4.4.

### 4.1 Computer Simulations of $q_n$ and $\rho_n^g$ in 3D-RSIM

In this article, we use the following computer simulations to calculate $q_n$ and $\rho_n^g$. Specifically: 1) Construct a three-dimensional simple cubic lattice with $400 \times 400 \times 400$ grid points; 2) For any lattice point, a random number ($r$) between 0 and 1 is firstly generated by the visual studio 2012 FORTRAN program, and there is no PS or a PS on the point if $r < \phi$ or $r \geq \phi$, respectively. Fig.1A illustrates the distribution of PSs in a y-z-plane in the simulation system when $\phi = 0.4$.

According to the construction method of PSSs in section 3, we construct PSSs in the above simulation systems (as illustrated in Fig.1B-C), and the resulting $q_n$ vs $n$ (Fig.2) can be described by the following exponential function,

$$q_n = q_0 e^{-n/n_0} \tag{4.1.1}$$

where $n_0$ is the average length of PSSs, and $q_0$ is the normalized constant (the normalized condition of $q_n$ used here is $\sum_{n=1}^{\infty} n q_n = 1$).

As shown in Fig.2, the simulated $q_n$ data fluctuates regularly around the fitted values of Eq.4.1.1 when $n$ is small, and the fluctuations decrease rapidly with increasing $n$. It is still unclear whether this question is derived from the construction method of PSs used in this paper or pseudo-random numbers.

The probability that a PS belongs to n-PSSs is $nq_n$, and $nq_n$ versus (vs) $n$ is shown in the inset of Fig.2. It can be seen that $nq_n$ appears as a single peak with the change of $n$, and the corresponding $n$ value ($n_p$) of the peak position decreases until $n_p = 1$ as $\phi$ increases (Table 1 in detail).

Here, the maximum value ($n_c$) of $n$ for numerical calculation is taken as where $nq_n$ is 1% of its maximum value and $n_c > n_p$, as shown in Table 1.

Table 1: $n_p$ and $n_c$ vs $\phi$

| $\phi$ | $n_p$ | $n_c$ |
|---|---|---|
| 0.10 | 17 | 132 |
| 0.20 | 12 | 90 |
| 0.30 | 9 | 70 |
| 0.40 | 7 | 54 |
| 0.50 | 5 | 42 |
| 0.60 | 4 | 31 |
| 0.70 | 3 | 21 |
| 0.80 | 2 | 14 |
| 0.85 | 1 | 11 |
| 0.90 | 1 | 8 |
| 0.95 | 1 | 5 |
| 0.99 | 1 | 3 |

For series of $\phi$ values, the resulting $n_0$ is shown in Fig.3. It can be seen that $n_0$ decreases with increasing $\phi$.

In this paper, the ring PSSs in 3D-RSIM are ignored (Fig.1C). The probability ($E_R$) that the PSs belong to the ring PSSs is equal to the number of all PSs in the ring PSSs dividing by the total number of PSs in the model. $E_R$ vs $\phi$ is shown in the inset of Fig.3, and the maximum value of $E_R$ is about 2.7%.

The simulated $\rho_n^g$ vs $g$ for series of $\phi$ and $n$ is shown in Fig.4 (the normalized condition of $\rho_n^g$ used is $\sum_{g=0}^{4n} \rho_n^g = 1$ in this paper). It is known that there is a single peak of $\rho_n^g$ with $g$ for all $\phi$ and $n$ (It could be imagined that, when $\phi \to 0$, $\rho_n^g$ is a Dirac $\delta$-function at $\frac{g}{n} = 4$). Representing $g$ corresponding to the maximum value of $\rho_n^g$ as $g_p^n$, it can be seen that: 1) There is a threshold at $\phi_s \approx 0.3$, and $g_p^n$ is nearly irrelevant to $n$ when $\phi \approx \phi_s$; 2) $g_p^n$ becomes larger or smaller as $n$ decreases for $\phi < \phi_s$ or $\phi > \phi_s$, respectively.

Fig.5 shows that, $\rho_n^g$ of $n = n_p$ changes with $g$ in 3D-RSIM for series of $\phi$, which indicates that: 1) There is another threshold at $\phi \approx 0.7$, and this value is near to the percolation threshold ($\phi_p = 0.69$) of 3D-RSIM [72] (Here, we consider this threshold is equal to $\phi_p$). When $\phi \geq \phi_p$, $g_p^n = 0$, and the $\rho_n^g$ peak becomes narrow with the increase of $\phi$; 2) For $\phi < \phi_p$, $g_p^n$ decreases, while $\rho_n^g$ peak widens as $\phi$ goes up.

## 4.2 Order Parameters, Static susceptibility and Specific Heat of 3D-RSIM

When the n-g-PSSs is in thermal equilibrium, the corresponding equilibrium value ($\eta_{ne}^g$) of $\eta_n^g$ is,

$$\eta_{ne}^g = \left[\frac{1}{nZ_n^g}\frac{\partial Z_n^g}{\partial \theta}\right]_{\theta=\theta_e} = \left[\gamma + \frac{1-\gamma^2}{nQ_n^g}\frac{\partial Q_n^g}{\partial \gamma}\right]_{\gamma=\gamma_e} \tag{4.2.1}$$

where $Z_n^g$ is the partition function of n-g-PSSs corresponding to $H_n^g$ (Appendix 3), also see Appendix 3 for $Q_n^g$, $\gamma_e \equiv \tanh(\theta_e)$, and $\theta_e \equiv \frac{\Theta_n^g}{T}\left[\left(1-\frac{1}{n}\right)\eta_{ne}^g + \frac{1}{n}\right]$.

From Eq.4.2.1, the static susceptibility ($\chi_s^{ng}$) of n-g-PSSs in thermal equilibrium is (Appendix 5),

$$\chi_s^{ng} = \frac{C_W}{N_0}\frac{n\aleph_n^g}{T-\aleph_n^g A_n^g} \tag{4.2.2}$$

Also see Appendix 5 for $\aleph_n^g$, $A_n^g \equiv \left(1-\frac{1}{n}\right)\Theta_n^g$, $C_W \equiv \frac{N_0\mu^2}{\varepsilon_0 k_B}$ is the Curie-Weiss constant, $N_0$ is the number of the lattice points per unit volume, and $\varepsilon_0$ is vacuum dielectric constant.

When n-g-PSSs is in thermal equilibrium, the average internal energy ($u_n^g$) and specific heat ($c_n^g$) per PS obtaining from $H_n^g$ of Eq2.3A is,

$$u_n^g = u_c^{ng} + u_f^{ng} \tag{4.2.3A}$$

$$u_c^{ng} = -\left[\frac{J}{Z_n^g}\frac{\partial Z_n^g}{\partial \varpi}\right]_{\theta=\theta_e} \tag{4.2.3B}$$

$$u_f^{ng} = -\frac{J}{2}\frac{g}{n}\left[\left(1-\frac{1}{n}\right)\eta_{nE}^g + \frac{1}{n}\right]\eta_{nE}^g \tag{4.2.3C}$$

$$c_n^g = \frac{\partial u_n^g}{\partial T} \tag{4.2.4}$$

Here, $u_c^{ng}$ and $u_f^{ng}$ are the average internal energies per PS, which correspond to the

intra- and inter-string interactions, respectively.

Fig.6 shows the variation of $\eta_{ne}^g$, $\chi_s^{ng}$, and $c_n^g$ with $T$ for $\frac{g}{n} = 3$ and series of $n$. Combining with Eq.3.2A, 2$^{nd}$ order CFPT (the peak value of $\chi_s^{ng}$ is infinite) can only occur in n-g-PSSs with $n \to \infty$. For finite $n$, n-g-PSSs undergoes a diffuse-ferroelectric-phase-transition (DFPT) ($\chi_s^{ng}$ (Fig.6b) and $c_n^g$ (Fig.6c) is a dispersion peak at a certain temperature), and the phase transition spreads to a wider temperature zone as $n$ decreases. In this article, we define the temperature corresponding to the maximum value of $-\frac{\partial \eta_{ne}^g}{\partial T}$ as the transition temperature ($T_p^{ng}$) of the DFPT of n-g-PSSs.

Fig.7 shows $\eta_{ne}^g$, $\chi_s^{ng}$, and $c_n^g$ vs $T$ for $n = 10$ and series of $g$, indicating that $T_p^{ng}$ of DFPT moves to low temperature as $g$ drops off (Fig.7b), as well as $\eta_{ne}^g$ and $c_n^g$ are not zero at temperature far higher than $T_p^{ng}$. $T_p^{ng} \equiv 0$ for n-g-PSSs with $g = 0$ (n-0-PSSs), giving that n-0-PSSs is PPSS; n-g-PSSs of $g \neq 0$ is FSS.

Physically, for finite $n$ and $g > 0$, the non-zero $\eta_{ne}^g$ value at temperature higher than $T_p^{ng}$ originates from that, the existence of the groups without PS makes the energy difference between the orientational configurations of all PSs becomes larger, leading to the probability of the ferroelectric configurations in 3D-RSIM being higher compared with 3D-IM [69].

The order parameter ($\eta$), spontaneous polarization ($P_s$), static susceptibility ($\chi_s^{ps}$), average internal energy ($u_{ps}$) and average specific heat ($c_{ps}$) per PS of 3D-RSIM are,

$$\eta = (1 - \phi) \sum_{n=1}^{\infty} \sum_{g=1}^{4n} n q_n \rho_n^g \eta_{nE}^g \tag{4.2.5A}$$

$$P_s = N_0 \mu \eta \tag{4.2.5B}$$

$$\chi_s^{ps} \approx (1 - \phi) \sum_{n=1}^{\infty} \sum_{g=0}^{4n} q_n \rho_n^g \chi_s^{ng} \tag{4.2.6}$$

$$u_{ps} = (1-\phi)\sum_{n=1}^{\infty}\sum_{g=0}^{4n} nq_n\rho_n^g u_n^g \qquad (4.2.7A)$$

$$c_{ps} = (1-\phi)\sum_{n=1}^{\infty}\sum_{g=0}^{4n} nq_n\rho_n^g c_n^g \qquad (4.2.7B)$$

In the calculation of macroscopic $\chi_s^{ps}$ by the microscopic $\chi_s^{ng}$, this paper uses an approximation similar to the parallel capacitance circuit (Eq.4.2.6).

For series of $\phi$, $P_s$, $\chi_s^{ps}$, and $c_{ps}$ vs $T$ are shown in Fig.8, and it can be seen that:

1) 3D-RSIM has a transition from $\eta = 0$ to $\eta \neq 0$ on cooling (Fig.8a), i.e. a ferroelectric phase transition occurs at a certain temperature.

2) As $\phi$ increases, $\frac{\eta(T\to 0)}{1-\phi}$ decreases (Fig.8a), indicating that only part of the PSs in 3D-RSIM undergoes ferroelectric phase transition. In fact, the n-0-PSSs in the model is the only PPSS as mentioned above, and the content ($R_P^0$) of PPSS in 3D-RSIM is,

$$R_P^0 = (1-\phi)\sum_{n=1}^{\infty} nq_n\rho_n^0 \qquad (4.2.9)$$

And the static susceptibility ($\chi_s^0$) of PPSS is,

$$\chi_s^0 \approx N_0(1-\phi)\sum_{n=1}^{\infty} q_n\rho_n^0 \chi_s^{n0} \qquad (4.2.10)$$

In 3D-RSIM, the transition from 0 to nonzero on cooling happens within a certain temperature range (Fig.8a), while $\chi_s^{ps}$ and $c_{ps}$ appear as diffuse peaks (Fig.8b-c). This is due to the spatial distribution of $T_p^{ng}$ of n-g-PSSs with different $n$ and $g$, so that the transition is an inhomogeneous-DFPT (IDFPT).

3) As $T$ decreases, there is an IDFPT when $\phi < 0.5$; two distinct IDFPTs (Fig.8b) near $\phi = \phi_p$, and which are called as low-temperature-IDFPT (LT-IDFPT) and high-temperature-IDFPT (HT-IDFPT), respectively; single IDFPT again for $\phi > 0.8$.

The detailed analysis shown in Fig.9 shows that LT-IDFPT and HT-IDFPT originate from n-g-PSSs of $g = 1$ (n-1-PSSs) and n-g-PSS of $g \geq 2$ (n-$2^+$-PSSs) in FSS, respectively. The order parameter ($\eta_1$), spontaneous polarization ($P_s^1$), static susceptibility ($\chi_s^1$), and average specific heat per PS ($c_{ps}^1$) of n-1-PSSs are,

$$\eta_1 \equiv (1-\phi) \sum_{n=1}^{\infty} n q_n \rho_n^1 \eta_{nE}^1 \qquad (4.2.10A)$$

$$P_s^1 \equiv N_0 \mu \eta_1 \qquad (4.2.10B)$$

$$\chi_s^1 \approx N_0(1-\phi) \sum_{n=1}^{\infty} q_n \rho_n^1 \chi_s^{n1} \qquad (4.2.11)$$

$$c_{ps}^1 = (1-\phi) \sum_{n=1}^{\infty} n q_n \rho_n^1 c_n^1 \qquad (4.2.12)$$

The order parameter ($\eta_{2+}$), spontaneous polarization ($P_s^{2+}$), static susceptibility ($\chi_s^{2+}$), and average specific heat per PS ($c_{ps}^{2+}$) of n-$2^+$-PSSs are,

$$\eta_{2+} = (1-\phi) \sum_{n=1}^{\infty} \sum_{g=2}^{4n} n q_n \rho_n^g \eta_{nE}^1 \qquad (4.2.13A)$$

$$P_s^{2+} = N_0 \mu \eta_{2+} \qquad (4.2.13B)$$

$$\chi_s^{2+} \approx N_0(1-\phi) \sum_{n=1}^{\infty} \sum_{g=2}^{4n} q_n \rho_n^g \chi_s^{ng} \qquad (4.2.14)$$

$$c_{ps}^{2+} = (1-\phi) \sum_{n=1}^{\infty} \sum_{g=2}^{4n} n q_n \rho_n^g c_n^g \qquad (4.2.15)$$

When $\phi$ is small, n-$2^+$-PSSs dominate FSS, so 3D-RSIM has a HT-IDFPT; FSS mainly contains n-1-PSSs for $\phi$ is large enough, which shows single LT-IDFPT; Near $\phi = \phi_p$, the contents of n-1-PSSs and n-$2^+$-PSSs are almost the same in FSS, and both the HT- and LT-IDFPT appear with $T$.

The contents of n-1-PSSs ($R_F^1$) and n-$2^+$-PSSs ($R_F^{2+}$) in 3D-RSIM are,

$$R_F^1 = (1-\phi) \sum_{n=1}^{\infty} n q_n \rho_n^1 \tag{4.2.16}$$

$$R_F^{2+} = (1-\phi) \sum_{n=1}^{\infty} \sum_{g=2}^{4n} n q_n \rho_n^g \tag{4.2.17}$$

In order to quantitatively describe the HT- and LT-IDFPT of 3D-RSIM, this paper defines: 1) The temperatures corresponding to the maximum values of $-\frac{d\eta_1}{dT}$ and $-\frac{d\eta_{2+}}{dT}$ being the phase transition temperatures of n-1-PSSs ($T_P^1$) and n-$2^+$-PSSs ($T_P^{2+}$); 2) To show the dispersion of the phase transitions, the diffuse temperatures ($T_d^1$ and $T_d^{2+}$) determined by $\frac{P_s^1(T_d^1)}{P_s^1(T_P^1)} \equiv \frac{1}{e}$ and $\frac{P_s^{2+}(T_d^{2+})}{P_s^{2+}(T_P^{2+})} \equiv \frac{1}{e}$; 3) The diffuse factors, $d_p^1 \equiv \frac{T_d^1 - T_P^1}{T_P^1}$ and $d_p^{2+} \equiv \frac{T_d^{2+} - T_P^{2+}}{T_P^{2+}}$, respectively.

The phase diagram of 3D-RSIM is shown in Fig.10, namely $T_P^1$, $T_P^{2+}$, $T_d^1$, $T_d^{2+}$, $d_p^1$, $d_p^{2+}$, $R_P^0$, $R_F^1$, $R_F^{2+}$, and $T_b$ (Burns temperature, see subsection 4.4 in detail) vs $\phi$, which indicate that:

1) As $\phi$ increases, $T_p^{2+}$ first decreases, but it remains almost unchanged after $\phi = \phi_p$; $T_d^{2+}$ first goes up slightly, then drops rapidly, and keeps as a constant after $\phi = 0.8$; $T_d^1$ first slowly, then drop rapidly, and increases slightly at the end; $d_p^{2+}$ shows a diffuse peak near $\phi = \phi_p$; $d_p^1$ decreases first, and then increases a little. $T_p^{2+}$ is always higher than $T_p^1$, and $T_p^1$ is almost irrelevant to $\phi$.

2) With increasing $\phi$, $R_P^0$ and $R_F^{2+}$, respectively, increases and decreases monotonically, with the maximum growth or drop rate near $\phi = 0.8$; $R_F^1$ shows a diffuse peak with the peak position near $\phi = 0.75$. In other words, 3D-RSIM has three subsystems: PPSS (n-0-PSSs), low-transition-temperature FSS (LTT-FSS, i.e. n-1-PSSs), and high-transition-temperature FSS (HTT-FSS, i.e. n-$2^+$-PSSs). The dominant subsystem is HTT-FSS when $\phi$ is small; PPSS, LTT-FSS, and HTT-FSS almost have the same contents near $\phi = \phi_p$; PPSS dominates the whole system when $\phi$ is large enough.

Along with the IDFPTs, there are at least three different kinds of interfaces in 3D-RSIM: 1) Phase boundaries: the interfaces between adjacent paraelectric and ferroelectric regions; 2) Sub-phase boundaries: the interfaces between adjacent ferroelectric regions of different $\eta_{ne}^g$ values; 3) Domain walls: the interfaces in ferroelectric regions with equal $\eta_{ne}^g$ values but opposite polarization directions. According to Imry-Ma theory [69], the ferroelectric regions in 3D-RSIM must become multi-domains due to the influence of the groups without PSs within and adjacent to the regions.

4.3 Complex Susceptibility of the Correlated Relaxation of PSs in 3D-RSIGM

In the vicinity of the transition or critical temperature of $2^{nd}$ order CFPT of typical ferroelectrics (such as TGS), not only there is the correlated relaxation of PSs (CR-PSs) (Referred as critical relaxation, phase transition relaxation [55,56]), but also it includes the relaxation of domain walls and other defects [76, 77].

In this section, the linear complex susceptibility ($\chi_n^{g*}$) of CR-PSs of n-g-PSSs is strictly calculated first based on Eq.3.2 (Appendix 6-9), and then by using $\chi_n^{g*}$, $q_n$, and $\rho_n^g$, the linear complex susceptibility ($\chi_{ps}^*$) of CR-PS in 3D-RSIGM is given according to the extended-Wagner-Approximation (EWA) [78].

The complex susceptibility (including the linear and high order) of n-g-PSSs is related to the change of $s_{nk}^g$ with time ($t$). Due to the interaction between PSs, the evolutions of $s_{nk}^g$ and $\zeta_{nk}^g$ ($k =1, \cdots n$) is interrelated (Appendix 6).

$\chi_n^{g*}$ is directly related to the sufficiently small deviation ($\delta_{nk}^g$) of $s_{nk}^g$ from its equilibrium value ($s_{nk}^{ge} = s_{nk}^g\big|_{\gamma=\gamma_e}$), i.e. $\delta_{nk}^g \equiv s_{nk}^g - s_{nk}^{ge}$. With $t$, $\delta_{nk}^g$ ($k = 1, \cdots n$) obey the following homogeneous linear equations (Appendix 8),

$$\frac{1}{v}\frac{d}{dt}\begin{bmatrix} \delta_{n1}^g \\ \vdots \\ \delta_{nn}^g \end{bmatrix} = -\begin{bmatrix} M_{1,1} & \cdots & M_{1,n} \\ \vdots & \ddots & \vdots \\ M_{n,1} & \cdots & M_{n,n} \end{bmatrix}\begin{bmatrix} \delta_{n1}^g \\ \vdots \\ \delta_{nn}^g \end{bmatrix} \tag{4.3.1}$$

where the square matrix $[M]_{n,n}$ is defined also in Appendix 8.

The physical meaning of the solution (Eqs.A8.1B, A8.2I and A8.3H in Appendix 8) of Eq.4.3.1 is that, the coupling relaxation of $\delta_{nk}^g$ ($k = 1, \cdots n$) is equivalent to $n$ mutually independent spatial-relaxation-modes (SRMs), which have different relaxation times ($\tau_{nk}^g$, $k = 1, \cdots n$) [79] and spatial distributions ($v_{nk}^{gi}$, $k, i = 1, \cdots n$). From short to long $\tau_{nk}^g$ ($k = 1, \cdots n$), all the SRMs are referred as 1st and nth SRM, respectively.

Fig.11 and Fig.12 show $v\tau_{nk}^g$ ($k = 1, \cdots n$) vs $T$ for series of $n$ and $g$. We can see that $v\tau_{nn}^g$ has a diffuse peak near $T_p^{ng}$, while $v\tau_{nk}^g$ ($k = 1, \cdots n-1$) has not the abnormal, and except $n = 1$, $v\tau_{nn}^g$ increases again with decreasing $T$, until diverges at $T = 0K$. In addition, the diffuse peak of $v\tau_{nn}^g$ gradually transits to a λ-type one as $n$ goes up (Fig.12a), and approximately $\tau_{nn}^g \sim \frac{1}{T-T_p^{ng}}$, which is consistent with the critical slowdown result of Mason theory [55]. Fig.12b indicates that, as long as $g \neq 0$, $v\tau_{nn}^g$ always have a diffuse peak near $T_p^{ng}$, and for all $g$ values, $v\tau_{nn}^g$ has the same divergent behavior at low temperatures.

Fig.13 and Fig.14 show $v_{nk}^{gi}$ ($k = 1, \cdots n$) of SRMs vs the PS position ($i = 1, \cdots n$) in n-g-PSSs when $n = 3$ and 4. Combining the results of $n = 2$ (Eq.A8.2I), we can conclude that the characteristic of $v_{nk}^{gi}$ is similar to that of standing waves. Similarly, this article also introduces the nodes of $v_{nk}^{gi}$ ($k, i = 1, \cdots n$), where $v_{nk}^{gi}$ is equal to zero, or two neighboring $v_{nk}^{gi}$ intersects at zero. Moreover, the nth, $\cdots$ 1st SRMs, respectively, have 0, $\cdots n-1$ nodes, and the larger the number of nodes, the more uneven the spatial distribution of SRM is; 2) For the nth, n-2th $\cdots$ SRMs whose node numbers are even, $v_{nk}^{gi} = v_{nk}^{gn-i-1}$ $(k = n, n-2 \cdots)$. For the n-1$^{th}$, n-3$^{th}$ $\cdots$ SRMs with odd number of nodes, there are $v_{nk}^{gi} = -v_{nk}^{gn-i-1}$ $(k = n-1, n-3 \cdots)$.

If the measured external field (electric field, stress, etc.) can be coupled with the SRMs, they can be detected. As shown in Appendix 9, the exact results of the real part ($\chi_n^{g\prime}$) and imaginary part ($\chi_n^{g\prime\prime}$) of $\chi_n^{g*}$ of n-g-PSSs with angular frequency (ω) is,

$$\chi_n^{g*} = \chi_n^{g\prime} - i_c \chi_n^{g\prime\prime} = \sum_{k=1}^{n} \frac{\Delta_{nk}^g}{1 + i_c \omega \tau_{nk}^g} \tag{4.3.2}$$

where $i_c$ is the imaginary unit, and $\Delta_{nk}^g$ is the dielectric relaxation strength of the k$^{th}$ SRM in n-g-PSSs,

$$\Delta_{nk}^g \equiv \frac{V_{nk}^{gn}}{\varepsilon_0 E_0} \sum_{i=1}^{n} v_{nk}^{gi} \tag{4.3.3}$$

Among them, $\frac{V_{nk}^{gn}}{\varepsilon_0 E_0}$ is the coupling strength of the k$^{th}$ SRM with a sufficiently small external electric field ($E_0$) in dielectric spectroscopy measurements. From Eq.4.3.3, $\Delta_{nk}^g = 0$ $(k = n-1, n-3 \cdots)$ due to that $v_{nk}^{gi} = -v_{nk}^{gn-i+1}$.

Fig.15 shows that, compared with $\Delta_{nn}^g$, the rest of $\Delta_{nk}^g$ $(k = 1, \cdots n-1)$ is much smaller, and by combining Eq.4.3.2 and A9.9, we get,

$$\chi_n^{g*} = \chi_n^{g\prime} - i_c \chi_n^{g\prime\prime} \approx \frac{\chi_s^{ng}}{1 + i_c \omega \tau_{nn}^g} \tag{4.3.4}$$

Fig.16 gives, when $\frac{g}{n} = 3$ and $U_B = 50J$, $\chi_n^{g\prime}$ and $\chi_n^{g\prime\prime}$ vs $T$, series of ω and $n$ calculated by Eq.4.3.4, which can be seen as follows: 1) When ω ≠ 0, $\chi_n^{g\prime}$ and $\chi_n^{g\prime\prime}$ have respectively a diffuse peak near $T_p^{ng}$ at the same time; 2) As ω increases, the peak temperatures of $\chi_n^{g\prime}$ and $\chi_n^{g\prime\prime}$, the low temperature side of $\chi_n^{g\prime}$ peak, and both of the high and low temperature sides of $\chi_n^{g\prime\prime}$ peak, all move toward high temperature, while the high temperature side of $\chi_n^{g\prime}$ peak is almost unchanged.

$\chi_n^{g*}$ in 3D-RSIGM has a distribution of both $n$ and $g$. However, there is still no rigorous theoretical method to calculate the complex susceptibility of heterogeneous systems on molecular scale. In this paper, EWA is used to calculate the real ($\chi_{ps}'$) and imaginary part ($\chi_{ps}''$) of the complex susceptibility ($\chi_{ps}^*$) of 3D-RSIGM, that is,

$$\chi_{ps}^* = \chi_{ps}' - i_c \chi_{ps}'' \approx N_0(1-\phi) \sum_{n=1}^{\infty} \sum_{g=0}^{4n} q_n \rho_n^g \chi_n^{g*} \tag{4.3.5}$$

In Eq.4.3.4 and Eq.4.3.5, if $\frac{\chi_s^{ng}}{n}$ is a constant independent of $n$ and $g$, that is the Wagner approximation (Corresponding to a parallel capacitor circuit) [78]. This is why we call Eq.4.3.5 as EWA in this paper.

For a system in which the relaxation units are randomly distributed in space, the Wagner approximation gives relatively accurate results only when the distribution of relaxation time is narrow. For 3D-RSIGM, since the distribution of $\tau_{nn}^g$ and $\chi_s^{ng}$ are wide (Fig.11, Fig.12, Fig.6b, and Fig.7b), the calculation of $\chi_{ps}^*$ by EWA (Eq.4.3.5) has relatively large errors.

Express the linear complex susceptibility of PPSS (n-0-PSSs), LTT-FSS (n-1-PSSs), and HTT-PSS (n-2$^+$-PSSs) in 3D-RSIGM as $\chi_{ps}^{0*}$, $\chi_{ps}^{1*}$, and $\chi_{ps}^{2+*}$, the corresponding real and imaginary parts are $\chi_{ps}^{0'}$, $\chi_{ps}^{0''}$, $\chi_{ps}^{1'}$, $\chi_{ps}^{1''}$, $\chi_{ps}^{2+'}$, and $\chi_{ps}^{2+''}$ respectively, it is obtained by Equation 4.3.5,

$$\chi_{ps}^{0*} = \chi_{ps}^{0'} - i_c \chi_{ps}^{0''} \approx N_0(1-\phi) \sum_{n=1}^{\infty} q_n \rho_n^0 \chi_n^{0*} \tag{4.3.6A}$$

$$\chi_{ps}^{1*} = \chi_{ps}^{1'} - i_c \chi_{ps}^{1''} \approx N_0(1-\phi) \sum_{n=1}^{\infty} q_n \rho_n^0 \chi_n^{1*} \tag{4.3.6B}$$

$$\chi_{ps}^{2+*} = \chi_{ps}^{2+'} - i_c \chi_{ps}^{2+''} \approx N_0(1-\phi) \sum_{n=1}^{\infty} \sum_{g=2}^{4n} q_n \rho_n^g \chi_n^{g*} \tag{4.3.6C}$$

Fig.17 shows, when $U_B = 50J$, $\phi$ =0.1、0.3、0.5 (HTT-FSS is the main in 3D-RSIM

from Fig.10b), the calculated $\chi'_{ps}$ and $\chi''_{ps}$ vs $T$ and $\omega$ by Eq.4.3.5, which indicates that: 1) Diffuse peaks of $\chi'_{ps}$ and $\chi''_{ps}$ appear near the IDFPT, and the peak temperature $(T_m)$ of $\chi'_{ps}$ is gradually approaching the peak temperature $(T_p^s)$ of $\chi_s^{ps}$, but the peak temperature of $\chi''_{ps}$ has no such limitation as $\omega$ increases; 2) The low temperature side of $\chi'_{ps}$ peak moves towards the high temperature, while the high temperature side changes little, and the peak height decreases with increasing $\omega$. Both the high and low temperature sides of $\chi''_{ps}$ peak shift to higher temperature, and the peak height has a tendency to goes up first and then drops as $\omega$ increases.

Arrehnius plot of $\omega$ vs $\frac{1}{T_m}$ ($10^{-8}\nu_0 - 10^{-3}\nu_0$) is shown in Fig.18, when $U_B = 50J$, $\phi = 0.1, 0.2, 0.3, 0.4$, and $0.5$, and it can be seen as follows: 1) Near critical slowdown of 2$^{nd}$ CFPT, i.e. $\omega \sim \frac{1}{T_m - T_p^s}$ for $\phi = 0.1$ and $0.2$; 2) Vogel-Fulcher type slowdown $(\ln(\omega) \sim \frac{1}{T_m - T_v}$, $T_v$ is Vogel temperature) [80] for $\phi = 0.3$ and $0.4$; 3) Arrehnius type slowdown for $\phi = 0.5$. Obviously, this is due to the distribution of local phase transitions.

Fig.19 shows that, when $U_B = 20J$ and $\phi = 0.8$ (PPSS, LTT-FSS, and HTT-PSS contents in 3D-RSIM are nearly same), $\chi'_{ps}$, $\chi^{0\prime}_{ps}$, $\chi^{1\prime}_{ps}$, and $\chi^{2+\prime}_{ps}$ vs $T$ and $\omega$ calculated by Eq.4.3.6, which indicates that: 1) Relaxation behaviors of $\chi^{1\prime}_{ps}$ and $\chi^{2+\prime}_{ps}$ are similar to the results of Figs.18a, c, and e; 2) Since the peak position of $\chi_s^0$ appears at 0K, the $\chi^{0\prime}_{ps}$ peak will move to low temperature until 0K with decreasing $\omega$.

It is worth noting that, except for $n = 1$, since $\nu\tau_n^{0n}$ rapidly increases at low temperature and diverges at 0K (Figs.12-13), PPSS is inevitably frozen in a certain temperature zone during a cooling process at a finite rate, and the freezing peak of $\chi_s^0$ occurs, so called PS glass transition happens. Considering the rapid increase of $R_P^0$ near $\phi_p$, $\phi_p$ is also defined as the characteristic concentration between IDFPT and PS glass systems in 3D-RSIGM (Fig.10).

In fact, except for $n = 1$, the average relaxation time of FSS also rapidly increases at low temperatures and diverges at 0K (Figs.13-14), so PS glass transitions will also occur during a cooling process with a finite rate, but the freeze peaks of $\chi_s^1$ and $\chi_s^{2+}$ are much weaker due to the influence of IDFPTs.

## 4.4 Burns Transformation in PRFEs

Currently, the interpretation to the Burns transformation of high temperature thermal strain ($s_{kl}^T$) and refractive index ($n_{kl}$, $k,l = 1,2,3$) in PRFEs is based on the macroscopic quadratic electro-strictive and Kerr (quadradic electro-optic) effects [80-86]. Burns et al. [80] propose that the transformation stems from the appearance of polar nanoregions (PNRs) during cooling.

Theoretically, the appearance of $P_s$ will inevitably lead to the deviations of $s_{kl}^T$ and $n_{kl}$ from the corresponding high temperature values. However, this does not rule out the possibility that other factors may cause the transformation. In this paper, based on the coupling between PSs and crystal lattice in 3D-RSIM, we give a new possible micro-mechanism of Burns transformation (Appendix 10-11 for details).

Considering the local-distortion (LD) of the crystal lattice and the change of local-electronic-clouds (LE) induced by the local interaction (LI) between the nearest-neighbor PSs in 3D-RSIM, which is abbreviated as LI-LD and LI-LE couplings respectively, and under the weak and linear LI-LD and LI-LE coupling approximations, the high temperature $s_{kl}^T$ and $n_{kl}$ of PRFEs are (Appendix 10-11),

$$s_{kl}^T - s_{kl}^0(T_r) \approx \alpha_{kl}(T - T_r) - c_{kl}\frac{u_{ps}}{J} \qquad (4.4.1)$$

$$n_{kl} - n_{kl}^0(T_r) \approx b_{kl}(T - T_r) - d_{kl}\frac{u_{ps}}{J} \qquad (4.4.2)$$

Among them, $c_{kl}$ and $d_{kl}$ are LI-LD and LI-LE coupling constants; $s_{kl}^0$ is the thermal strain caused by the non-harmonic part of the interaction constructing crystal lattice; $n_{kl}^0$

is the refractive index independent of LI-LE; $\alpha_{kl}$ and $b_{kl}$ are respectively the high-temperature thermal expansion and thermo-optic coefficients caused by the non-harmonic interaction; $T_r$ is a reference temperature.

When the temperature is higher, $u_{ps} \to 0$, so $s_{kl}^T - s_{kl}^0(T_r) \approx \alpha_{kl}(T - T_r)$ and $n_{kl} - n_{kl}^0(T_r) \approx b_{kl}(T - T_r)$. As temperature decreases, $|u_{ps}|$ increases (Eq.4.2.3, Fig.8c), resulting in the deviation of $s_{kl}^T$ and $n_{kl}$ from their linear behaviors of high temperature, shown in Fig.20. In addition, the Burns temperature $(T_b)$ increases as $c_{kl}$ and $|d_{kl}|$ increases. Therefore, according to the above results, the Burns transformation in PRFEs is a crossover of the nonharmonic interaction to LI-LD and LI-LE couplings.

## 5. Discussions

This section mainly includes the comparison of the model predictions with Monte Carlo simulations and PRFE experiments is given in subsections 5.1 and 5.2 respectively. The potential future study is presented in subsection 5.3.

### 5.1 Comparison of Model Predictions with Monte Carlo Simulations

At present, Monte Carlo simulations in 3D- and 2D-RSIM only give the results of specific heat and spontaneous polarization when $\phi < \phi_p$ [50,51]. The main conclusions are as follows: 1) The system has a diffuse ferroelectric/ferromagnetic phase transition; 2) The transition dispersity is enhanced and the transition temperature decreases with the increase of $\phi$. This is consistent with the MF-PSS predictions (Fig.8a and c). It should be noted that the phase transition temperature given by the MF-PSS is slightly higher than that of the simulations, which is the commonality of the mean field of Weiss type.

### 5.2 Comparison of Model Results with PRFE Experiments

The experimental results of $P_s$ [86-89] and domain structures [90] in $S_xB_{1-x}Nb_2O_3$,

PM$_{1/3}$N$_{2/3}$, BZ$_x$T$_{1-x}$, etc. show that $P_s \neq 0$ below the RFPT, indicating that RFPT is a ferroelectric phase transition. Also, due to the diffusive peak of $-\frac{\partial P_s}{\partial T}$ with $T$, rather than the λ-type peak of CFPT, RFPT is a diffuse phase transition. The domain structures of PRFEs shows that the microscopic regions of ferroelectric and paraelectric phases coexist, and $P_s$ is spatially inhomogeneous, so RFPT is also an inhomogeneous phase transition. Therefore, RFPT is an IDFPT, which is consistent with the model results (Fig.8a). It is conceivable that RISF and RIEF will further enhance the dispersity of the IDFPT.

Specific heat experiments [91-94] of PbZn$_{1/3}$Nb$_{2/3}$O$_3$, Sr$_x$Ba$_{1-x}$NbO$_6$, PM$_{1/3}$N$_{2/3}$, and PbMg$_{1/3}$Ta$_{2/3}$O$_3$ show that, the specific heat peak corresponding to RFPT appears in PRFEs, which is also consistent with the model results as shown in Fig.8c. According to Refs. [91, 94], the peak heights in PbZn$_{1/3}$Nb$_{2/3}$O$_3$, PM$_{1/3}$N$_{2/3}$, and PbMg$_{1/3}$Ta$_{2/3}$O$_3$ single crystal samples are approximately 7, 5, and 4 JK$^{-1}$mol$^{-1}$, respectively; when $\phi = \frac{1}{3}$, the result of 3D-RSIM given by the MF-PSSs is about 5.6 JK$^{-1}$mol$^{-1}$ (Fig.8c). So, the theoretical and experimental results are also almost equal to each other.

For 3D-RSIGM of $\phi < \phi_p$, the phase boundaries, sub-phase boundaries, and domain walls generated during the IDFPT will inevitably contribute significantly to the complex susceptibility below the transition temperature. Here, they are collectively referred as $\chi_b^*$. Since the relaxations of the boundaries and domain walls are all derived from the overall movement of large number PSs, the average relaxation time of $\chi_b^*$ is longer than that of $\chi_{ps}^*$. Therefore, the main contributions of the complex susceptibility of 3D-RSIGM of $\phi < \phi_p$ are, respectively, $\chi_b^*$ or $\chi_{ps}^*$ when $T < T_p^{2+}$ or $T > T_p^{2+}$.

For the typical PRFE PM$_{1/3}$N$_{2/3}$, the specific heat experiments [91,94] show that $T_p^{2+}(\text{PM}_{1/3}\text{N}_{2/3}) \approx 300$K, and the frequency when $T_m \approx T_p^{2+}(\text{PM}_{1/3}\text{N}_{2/3})$ is about 200 MHz according to the susceptibility data [31]. Therefore, according to the model predictions of this paper, in PM$_{1/3}$N$_{2/3}$, the complex susceptibility for ω higher than 200π

MHz is mainly $\chi_{ps}^*$. The experimental results of PM$_{1/3}$N$_{2/3}$ single crystals (Fig.1 of Ref.31]) show that, with increasing ω, the real part peak of the complex susceptibility moves to high temperature that gives the Vogel-Fulcher relation of $T_m$ with ω, meanwhile its peak height decreases, and the imaginary part peak also shifts to high temperature, but its peak height goes up first and then drops. This is consistent with the behaviors of $\chi_{ps}^*$ and the Vogel-Fulcher type slowdown of $T_m$ with ω predicted by the present model (Fig.17b, Fig.17d, and Fig.18). Because the measured data of $T_m$ vs ω by different research groups have significant differences in Sr$_x$Ba$_{1-x}$Nb$_2$O$_6$ [35,36], they are not compared with the model predictions here.

Since x can be continuously adjusted from 0 to 1, BZ$_x$T$_{1-x}$ is an ideal system to verify the model of this paper. However, due to the limitation of single crystal growth technology, large-size and high-quality single crystals with $x > 0.2$ cannot be grown, which has affected the measurement of some physical parameters (Especially high frequency complex susceptibility) to some extent [37,45]. In addition, because 3D-RSIGM is used for the first time to study RFPT in detail, the relationship between the key model parameters (ϕ, J, μ, and $U_B$) and x is not yet clear, there is a certain difficulty in comparison with the experimental results. However, taking into account the similarity of the real part of low and high frequency complex susceptibility of typical PRFE PM$_{1/3}$N$_{2/3}$ [31], the analogy of $T_m$ vs x given by the real part of the low frequency complex susceptibility of BZ$_x$T$_{1-x}$ with the corresponding model result of this paper has a certain degree of rationality.

The experimental results of the real part of the low frequency (100Hz-500kHz) complex susceptibility of BZ$_x$T$_{1-x}$ [37] show that: 1) When x increases from 0.15 to 0.75, $T_m$ first decreases linearly with x, and then is almost constant after $x \sim 0.5$ (Fig.11 of Ref.37) This is very similar to the variation trend of $T_p^{2+}$ with ϕ predicted by the present model, except that $\phi_p$ corresponds to $x \approx 0.45$. This may be due to the RISF produced by Zr$^{4+}$, which annihilates some of the permanent dipoles; 2) When x increases from 0.35 to 0.75, the change of $T_m$ vs ω varies from the Vogel-Fulcher to Arrehnius type slowdown (Fig.8 of Ref.37). This is also consistent with the results of the model in this paper (Fig.18).

The existing experimental results show that there are successively intermediate temperature ($T^*$) [84-85] and $T_b$ [80-83] above the RFPT. According to the current view, $T_b$ is the temperature at which PNRs is generated on cooling, and $T^*$ is the temperature where PNRs changes from the high temperature dynamic to low temperature static. The model in this paper predicts the existence of $T_d^{2+}$ and $T_b$ (Fig.10a and Fig.20) above $T_p^{2+}$ of the IDFPT. According to $T_p^{2+}(PM_{1/3}N_{2/3}) \approx 300K$ [91,94], the model predicts $T_d^{2+}(PM_{1/3}N_{2/3}) \approx 380K$ (Fig.10a), which is very close to $T^* \approx 400K$ given by the neutron scattering experiment [84,85]. In other words, this article may give a new possible mechanism of $T^*$.

In addition, according to the model results in this paper, at the temperatures of $T_b \approx 620K$ [5,80-82], $\frac{T_d^{2+}+T_b}{2} \approx 500K$, and $T_d^{2+} \approx 380K$ in PM$_{1/3}$N$_{2/3}$, $\eta \approx$ 0.04, 0.06, and 0.14, and the ratio of the internal energy ($u_{ps}^f$) contributed by $\eta$ to $u_{ps}$, i.e. $\frac{u_{ps}^f}{u_{ps}} \approx$ 7%, 9%, and 18%, respectively. According to Eq.4.4.1 and Eq.4.4.2, the contribution of the spontaneous polarization to the thermal strain and refractive index is secondary between $T_d^{2+}$ and $T_b$ in PM$_{1/3}$N$_{2/3}$, and the main comes from the correlation between PSs. This result also avoids the too large $P_s$ obtained by the quadradic electrostrictive and Kerr effects (Eqs.A10.3 and A11.3), compared to the data of hysteresis and pyroelectric measurements [5]. In other words, above $T_d^{2+}$ in PRFEs, PNRs with excessive $P_s$ may not exist.

## 5.3 Potential Future Study

Closely related to the model of this paper, the potential future studies are: 1) To explore new methods that are more accurate than the EWA to calculate $\chi_{ps}^*$ of 3D-RSIGM, so that quantitative comparison with the corresponding experiments of PRFEs can be performed;

2) To introduce RISF and RIEF into 3D-RSIGM, and the solution of the corresponding models; 3) The relationship of $\phi$, $J$, $\mu$, and $U_B$ with x in variable component PRFEs (such as $BZ_xT_{1-x}$, and $Sr_xBa_{1-x}Nb_2O_6$); 4) To generalize the MF-PSSs for solving isotropic 3D-RSIGM to anisotropic cases, so that the anisotropic PRFEs of single-axis tungsten bronze and layer Aurivillius structures can be described; 5) To generalize the model predictions of this paper to the corresponding ferromagnetic systems, in particular spin-glasses; 6) Strong coupling of PSs with crystal lattice and possible structural phase transitions, etc.

## 6. Conclusion

In this paper, 3D-RSIGM of PSs is applied for the first time to study the RFPT of PRFEs in detail. To solve this model, we proposed a new method, MF-PSSs, which includes the effect of the interfaces between the groups with or without PSs. We find:

1) 3D-RSIM is a mixed system consisting of PPSS, LTT-FSS, and HTT-FSS. The contents of these three subsystems change with $\phi$.

2) When $\phi < \phi_p$, HTT-FSS is the dominant of the model. On cooling, the model system undergoes an IDFPT, and as $\phi$ increases, the relationship between $T_m$ and $\omega$ shows a crossover from the critical, gradually to the Vogel-Fulcher, and to the Arrehnius type slowdown.

3) When $\phi = \phi_p$, the three subsystems have comparable contents. However, when $\phi > \phi_p$, PPSS becomes the dominating subsystem. On cooling, the spontaneous polarization and specific heat of the whole system show an IDFPT, but $T_m$ shifts to lower temperature with decreasing $\omega$ until 0K. Since the average relaxation time of PPSS rapidly increases at lower temperature until it diverges at 0K, during a cooling process of a limited rate, PPSS must be frozen at a certain temperature, i.e. PS glass transition occurs. In short, with the increase of $\phi$, 3D-RSIGM gradually evolves from the IDFPT system to PS glass, and $\phi_p$ is the characteristic concentration of this evolution.

Moreover, based on the coupling between PSs and crystal lattice, we also give a new possible mechanism of Burns transformation.
4) The model predictions in this paper is consistent with the corresponding Monte Carlo simulation and typical PRFE experimental results.

## Appendix 1: Glauber Transition Probability for 3D-RSIM

By the detailed balance condition of the Hamiltonian (Eq.2.1A), it is obtained that,

$$w(\sigma_i)\exp\left(\frac{\sigma_i H_i}{k_B T}\right) = w(-\sigma_i)\exp\left(-\frac{\sigma_i H_i}{k_B T}\right) \tag{A1.1}$$

Similar to Glauber's choice of the transition probability of one-dimensional Ising model [52], this paper chooses the $w(\sigma_i)$ as,

$$w(\sigma_i) = \frac{v}{2}\left[1 - \sigma_i \tanh\left(\frac{H_i}{k_B T}\right)\right] \tag{A1.2}$$

For the Hamiltonian of Equation 3.2A, the local field ($H_{ni}^g$) of $\sigma_i$ in n-g-PSSs is,

$$H_{ni}^g \equiv J\left\{\sigma_{i-1} + \sigma_{i+1} + \frac{g}{n}\left[\left(1 - \frac{1}{n}\right)\eta_n^g + \frac{1}{n}\right]\right\} \tag{A1.3}$$

where $i = 1, \cdots n$, $\sigma_0 = \sigma_{n+1} = 0$, and,

$$\begin{cases}\exp\left(-\frac{\sigma_1 H_{n1}^g}{k_B T}\right) \propto 1 - \gamma\sigma_1 + \alpha(\gamma - \sigma_1)\sigma_2 \\ \exp\left(-\frac{\sigma_n H_{nn}^g}{k_B T}\right) \propto 1 - \gamma\sigma_n + \alpha(\gamma - \sigma_n)\sigma_{n-1} \\ \exp\left(-\frac{\sigma_i H_{ni}^g}{k_B T}\right) \propto 1 - \gamma\sigma_i + \beta(\gamma - \sigma_i)(\sigma_{i-1} + \sigma_{i+1})\end{cases}$$

Also similar to the Glauber's choice [52], we get:

For $n \leq 2$,

$$w(\sigma_i) = \frac{v}{2}[1 - \gamma\sigma_i + \alpha(\gamma - \sigma_i)(\sigma_{i-1} + \sigma_{i+1})] \tag{A1.4A}$$

For $n \geq 3$,

$$\begin{cases}w(\sigma_1) = \frac{v}{2}[1 - \gamma\sigma_1 + \alpha(\gamma - \sigma_1)\sigma_2] \\ w(\sigma_n) = \frac{v}{2}[1 - \gamma\sigma_n + \alpha(\gamma - \sigma_n)\sigma_{n-1}] \\ w(\sigma_i) = \frac{v}{2}[1 - \gamma\sigma_i + \beta(\gamma - \sigma_i)(\sigma_{i-1} + \sigma_{i+1})]\end{cases} \tag{A1.4B}$$

where $i = 2, \cdots n - 1$.

## Appendix 2: Weiss Mean-Field Considering Interface Effect

Ferdinand and Fisher first [69] strictly calculated the phase transition specific heat of finite two-dimensional Ising model containing $n \times n$ spins. By comparing with the infinite system, it was found that the boundary effect of the finite system caused the specific heat peak to disperse, and the peak temperature decreases, as well as the influence factor is approximately proportional to $\frac{1}{n}$. This can be understood as that the system energy and interface energy of two-dimensional Ising model are respectively proportional to $n^2$ and $n$, so the boundary effect factor is approximately proportional to the ratio of the interface energy to the system energy, i.e. $\frac{1}{n}$. For finite three-dimensional Ising model, it can also be obtained that the factor is approximately proportional to $\frac{1}{n}$.

For 3D-RSIM, the interfaces between the groups with or without PSs will inevitably lead to the interface effect. This paper also takes that the interface effect factor is approximately proportional to $\frac{1}{n}$.

At present, there are two schemes to introduce the interface effect in the Weiss mean-field: 1) Introduce an internal field unrelated to the order parameter ($\eta$); 2) Modify the Weiss mean-field proportional to the fractional mean-field ($\eta^f$), where $0 < f < 1$.

This paper chooses the first scheme, and the internal field is taken as $-J\frac{g}{n}\frac{f_b}{n}$, where $f_b$ is the geometric factor of the interfaces ($f_b = 1$ is used herein). In addition, the interfacial effect corrects the Weiss mean-field to $-J\frac{g}{n}\left(1 - \frac{1}{n}\right)\eta_n^g$.

## Appendix 3: Calculation of $Z_n^g$ and $Q_n^g$

The partition function ($Z_n^g$) of n-g-PSSs corresponding to $H_n^g$ is,

$$Z_n^g \equiv \sum_{\sigma_1 \cdots \sigma_n} \exp\left[\varpi \sum_{i=1}^{n-1} \sigma_i \sigma_{i+1} + \theta \sum_{j=1}^{n} \sigma_j\right]$$

From $\exp(\varpi \sigma_i \sigma_{i+1}) = (1 + \alpha \sigma_i \sigma_{i+1})\cosh(\varpi)$ and $\exp(\theta \sigma_j) = (1 + \gamma \sigma_j)\cosh(\theta)$, we obtain,

$$Z_n^g = Q_n^g \cosh^{n-1}(\varpi) \cosh^n(\theta) \tag{A3.1}$$

where,

$$Q_n^g \equiv \sum_{\sigma_1 \cdots \sigma_n} \prod_{i=1}^{n-1}(1 + \alpha \sigma_i \sigma_{i+1}) \prod_{j=1}^{n}(1 + \gamma \sigma_j) \tag{A3.2}$$

Introduce a variable,

$$Y_n^g \equiv \sum_{\sigma_1 \cdots \sigma_n} \prod_{i=1}^{n-1}(1 + \alpha \sigma_i \sigma_{i+1}) \prod_{j=1}^{n}(1 + \gamma \sigma_j) \sigma_n \tag{A3.3}$$

Obviously, $Q_1^g = 2$ and $Y_1^g = 2\gamma$.

The following recurrence formulas are derived from Eqs.A3.2-3,

$$\begin{cases} Q_n^g = 2(Q_{n-1}^g + \alpha \gamma Y_{n-1}^g) \\ Y_n^g = 2(\gamma Q_{n-1}^g + \alpha Y_{n-1}^g) \end{cases} \tag{A3.4}$$

Let,

$$A_n^g \equiv Q_n^g + B Y_n^g \tag{A3.5}$$

to make,

$$A_n^g = D A_{n-1}^g \tag{A3.6}$$

where $B$ and $D$ are pending constants.

From Eqs.A3.4-6, it is obtained that,

$$A_n^g = \frac{2}{D}\left[(1 + \gamma B)Q_{n-1}^g + \alpha(\gamma + B)Y_{n-1}^g\right] \tag{A3.7}$$

and by the comparison of this equation with Eqs.A3.5-6, we get,

$$\begin{cases} \dfrac{2(1 + \gamma B)}{D} = 1 \\ \dfrac{2\alpha(\gamma + B)}{D} = B \end{cases} \tag{A3.8}$$

From this equation, two values of $B$ and $D$ are,

$$\begin{cases} B_1 = \dfrac{\alpha - 1 + \Omega}{2\gamma} \\ B_2 = \dfrac{\alpha - 1 - \Omega}{2\gamma} \end{cases} \tag{A3.9}$$

$$\begin{cases} D_1 = 2(1 + \gamma B_1) = 1 + \alpha + \Omega \\ D_2 = 2(1 + \gamma B_2) = 1 + \alpha - \Omega \end{cases} \tag{A3.10}$$

where $\Omega \equiv \sqrt{(1 - \alpha)^2 + 4\alpha\gamma^2}$.

This gives two values of $A_n^g$,

$$\begin{cases} A_{n1}^g = D_1^{n-1}(Q_1^g + B_1 Y_1^g) = D_1^n \\ A_{n2}^g = D_2^{n-1}(Q_1^g + B_2 Y_1^g) = D_2^n \end{cases} \tag{A3.11}$$

By Eqs.A3.5 and A3.11, it is obtained,

$$\begin{cases} Q_n^g + B_1 Y_n^g = D_1^n \\ Q_n^g + B_2 Y_n^g = D_2^n \end{cases} \tag{A3.12}$$

Therefore,

$$\begin{cases} Q_n^g = G_1 D_2^n + G_2 D_1^n \\ Y_n^g = \dfrac{\gamma}{\Omega}(D_1^n - D_2^n) \end{cases} \tag{A3.13}$$

where $G_1 \equiv \dfrac{B_1}{B_1 - B_2} = \dfrac{1}{2} - \dfrac{1-\alpha}{2\Omega}$ and $G_2 \equiv \dfrac{B_2}{B_2 - B_1} = \dfrac{1}{2} + \dfrac{1-\alpha}{2\Omega}$.

## Appendix 4: Calculation of $s_{nk}^g$

According to $H_n^g$ (Eq.3.2A), the expectation value ($s_{nk}^g$) of $\sigma_k$ in n-g-PSSs is,

$$s_{nk}^g \equiv \frac{1}{Z_n^g} \sum_{\sigma_1 \cdots \sigma_n} \sigma_k \exp\left[\varpi \sum_{i=1}^{n-1} \sigma_i \sigma_{i+1} + \theta \sum_{j=1}^{n} \sigma_j\right] \tag{A4.1}$$

and,

$$s_{nk}^g = \frac{1}{Q_n^g} \sum_{\sigma_1 \cdots \sigma_n} \prod_{i=1}^{n-1}\prod_{j=1}^{n}(1 + \alpha \sigma_i \sigma_{i+1})(1 + \gamma \sigma_j)\, \sigma_k \tag{A4.2}$$

By $(1 + \alpha \sigma_k \sigma_{k+1})\sigma_k = \sigma_k + \alpha \sigma_{k+1}$, we get,

$$s_{nk}^g = \frac{1}{Q_n^g} \left[ \sum_{\sigma_1\cdots\sigma_k} \prod_{i=1}^{k-1} \prod_{j=1}^{k} (1+\alpha\sigma_i\sigma_{i+1})(1+\gamma\sigma_j)\sigma_k \sum_{\sigma_{k+1}\cdots\sigma_n} \prod_{i=k+1}^{n-1} \prod_{j=k+1}^{n} (1 \right.$$

$$+ \alpha\sigma_i\sigma_{i+1})(1+\gamma\sigma_j)$$

$$+ \alpha \sum_{\sigma_1\cdots\sigma_k} \prod_{i=1}^{k-1} \prod_{j=1}^{k} (1+\alpha\sigma_i\sigma_{i+1})(1+\gamma\sigma_j) \sum_{\sigma_{k+1}\cdots\sigma_n} \prod_{i=k+1}^{n-1} \prod_{j=k+1}^{n} (1$$

$$\left. + \alpha\sigma_i\sigma_{i+1})(1+\gamma\sigma_j)\sigma_{k+1} \right]$$

that is,

$$s_{nk}^g = \frac{1}{Q_n^g}(Y_k^g Q_{n-k}^g + \alpha Q_k^g Y_{n-k}^g) \tag{A4.3}$$

and,

$$s_{nk}^g = s_{nn-k+1}^g \tag{A4.4}$$

## Appendix 5: Calculation of $\chi_s^{ng}$

Along the direction of PSs in 3D-RSIGM, the small enough external electric field ($E$) is loaded, the Hamiltonian of n-g-PSSs becomes,

$$H_{nE}^g = -J\sum_{i=1}^{n-1}\sigma_i\sigma_{i+1} - \left\{J\frac{g}{n}\left[\left(1-\frac{1}{n}\right)\eta_n^{gE} + \frac{1}{n}\right] + \mu E\right\}\sum_{j=1}^{n}\sigma_j \tag{A5.1}$$

and the corresponding partition function ($Z_{nE}^g$) and order parameter ($\eta_n^{gE}$) are respectively,

$$Z_{nE}^g \equiv \sum_{\sigma_1\cdots\sigma_n} \exp\left[\varpi\sum_{i=1}^{n-1}\sigma_i\sigma_{i+1} + \theta^E\sum_{j=1}^{n}\sigma_j\right] \tag{A5.2A}$$

$$\eta_n^{gE} = \frac{1}{nZ_{nE}^g}\frac{\partial Z_{nE}^g}{\partial \theta^E} \tag{A5.2B}$$

where $\theta^E \equiv \theta + \frac{\mu E}{k_B T}$. Since $Z_{nE}^g$ is an even function about $E$, $Z_{nE}^g = Z_n^g + O(E^2)$ as $E \to 0$。

By Eqs.A5.1-2, the static susceptibility ($\chi_s^{ng}$) of n-g-PSSs in thermal equilibrium is,

$$\chi_s^{ng} \equiv \frac{n\mu}{\varepsilon_0} \frac{\partial \eta_n^{gE}}{\partial E} = \frac{n\mu}{\varepsilon_0} \frac{\partial \eta_n^{gE}}{\partial \theta^E} \frac{\partial \theta^E}{\partial E} = \frac{\partial \eta_n^{gE}}{\partial \theta^E} \left[ \frac{n\mu}{\varepsilon_0} \frac{\partial \eta_n^{gE}}{\partial E} \frac{A_n^g}{T} + \frac{n}{N_0} \frac{C_w}{T} \right]$$

$$= \aleph_n^g \left( \frac{A_n^g}{T} \chi_s^{n,g} + \frac{n}{N_0} \frac{C_w}{T} \right)$$

and, we get,

$$\chi_s^{ng} = \frac{nC_w}{N_0} \frac{\aleph_n^g}{T - \aleph_n^g A_n^g} \tag{A5.3}$$

where $\aleph_n^g \equiv \left. \frac{\partial \eta_n^{gE}}{\partial \theta^E} \right|_{E \to 0} = \frac{\partial \eta_{ne}^g}{\partial \theta_e} = \left\{ 1 - \gamma^2 + \frac{1}{nQ_n^g} \left[ \frac{\partial^2 Q_n^g}{\partial \theta^2} - \frac{1}{Q_n^g} \left( \frac{\partial Q_n^g}{\partial \theta} \right)^2 \right] \right\}_{\gamma=\gamma_e, \theta=\theta_e}$, and $A_n^g \equiv \left( 1 - \frac{1}{n} \right) \Theta_n^g$.

## Appendix 6: Derivation of the Relaxation Equation for $s_{nk}^g$

As shown in Equs3.2B-D, the Glauber transition probability of n-PSSs and infinite PSS chains are slightly different. For the sake of clarity, the following derivation of relaxation equations for $s_{nk}^g$ is given here.

In n-g-PSSs, the probability that $n$ PSs take a specific value $(\sigma_1, \cdots \sigma_n)$ is,

$$p(\sigma_1, \cdots \sigma_n) = \frac{1}{Z_n^g} \exp\left[ \varpi \sum_{i=1}^{n-1} \sigma_i \sigma_{i+1} + \theta \sum_{j=1}^{n} \sigma_j \right] \tag{A6.1}$$

Then,

$$\frac{dp(\sigma_1, \cdots \sigma_n)}{dt} = \sum_i^n w(-\sigma_i) p(\sigma_1, \cdots -\sigma_i, \cdots \sigma_n)$$

$$- \sum_j^n w(\sigma_i) p(\sigma_1, \cdots \sigma_i, \cdots \sigma_n) \tag{A6.2}$$

For,

$$s_{nk}^g = \sum_{\sigma_1,\cdots\sigma_n} \sigma_k p(\sigma_1,\cdots\sigma_n) \tag{A6.3}$$

Multiply both sides of Eq.A6.2 by $\sigma_k$ and calculate the sum of $\sigma_1,\cdots\sigma_n$, then obtain,

$$\frac{ds_{nk}^g}{dt} = \sum_{\sigma_1,\cdots\sigma_n} \sigma_k \left[\sum_{i}^{n} w(-\sigma_i)p(\sigma_1,\cdots-\sigma_i,\cdots\sigma_n) \right.$$
$$\left. -\sum_{j}^{n} w(\sigma_i)p(\sigma_1,\cdots\sigma_i,\cdots\sigma_n)\right] \tag{A6.4}$$

In the right side of Eq.A6.4, the term of $i = k$ is equal to,

$$\sum_{\sigma_1,\cdots\sigma_n} \sigma_k[w(-\sigma_k)p(\sigma_1,\cdots-\sigma_k,\cdots\sigma_n) - w(\sigma_k)p(\sigma_1,\cdots\sigma_k,\cdots\sigma_n)]$$

$$= 2\sum_{\sigma_1,\cdots\sigma_n} \sigma_k w(\sigma_k)p(\sigma_1,\cdots\sigma_k,\cdots\sigma_n) \tag{A6.5}$$

and by,

$$\sum_{\sigma_i} [w(-\sigma_i)p(\sigma_1,\cdots-\sigma_i,\cdots\sigma_n) - w(\sigma_i)p(\sigma_1,\cdots\sigma_i,\cdots\sigma_n)] = 0$$

we obtain that, in the right side of Eq.A6.4, the term of $i \neq k$ is equal zero. So,

$$\frac{ds_{nk}^g}{dt} = -2\sum_{\sigma_1,\cdots\sigma_n} \sigma_k w(\sigma_k)p(\sigma_1,\cdots\sigma_n) \tag{A6.6}$$

Substituting the expression for $w(\sigma_k)$ (Eqs.3.2B-D), the relaxation equations for $s_{nk}^g$ are,

for $n = 1$,
$$\frac{1}{v}\frac{ds_{11}^g}{dt} = -s_{11}^g + \gamma \tag{A6.7A}$$

for $n = 2$,
$$\begin{cases} \dfrac{1}{v}\dfrac{ds_{21}^g}{dt} = -s_{21}^g + \alpha s_{22}^g + \gamma(1 - \alpha\zeta_{21}^g) \\ \dfrac{1}{v}\dfrac{ds_{22}^g}{dt} = -s_{22}^g + \alpha s_{21}^g + \gamma(1 - \alpha\zeta_{21}^g) \end{cases} \tag{A6.7B}$$

for $n \geq 3$,

$$\begin{cases} \dfrac{1}{\nu}\dfrac{ds_{n1}^g}{dt} = -s_{n1}^g + \alpha s_{n2}^g + \gamma(1-\alpha\zeta_{n1}^g) \\ \dfrac{1}{\nu}\dfrac{ds_{nn}^g}{dt} = -s_{n1}^g + \alpha s_{nn-1}^g + \gamma(1-\alpha\zeta_{nn-1}^g) \\ \dfrac{1}{\nu}\dfrac{ds_{nk}^g}{dt} = -s_{nk}^g + \beta(s_{nk-1}^g + s_{nk+1}^g) + \gamma[1-\beta(\zeta_{nk-1}^g + \zeta_{nk+1}^g)] \end{cases} \quad (A6.7C)$$

where $k = 2, \cdots n-1$。

## Appendix 7: Calculation of $\zeta_{nk}^g$

In n-g-PSSs, the correlation function ($\zeta_{nk}^g$) between the nearest neighbor $k^{th}$ and $k+1^{th}$ PSs, i.e. the expectation value of $\sigma_k \sigma_{k+1}$ is,

$$\zeta_{nk}^g \equiv \frac{1}{Z_n^g} \sum_{\sigma_1 \cdots \sigma_n} \sigma_k \sigma_{k+1} \exp\left[\varpi \sum_{i=1}^{n-1} \sigma_i \sigma_{i+1} + \theta \sum_{j=1}^{n} \sigma_j\right]$$

$$= \frac{1}{Q_n^g} \sum_{\sigma_1 \cdots \sigma_n} \prod_{i=1}^{n-1}(1+\alpha\sigma_i\sigma_{i+1}) \prod_{j=1}^{n}(1+\gamma\sigma_j) \sigma_k \sigma_{k+1}$$

By $(1+\alpha\sigma_k\sigma_{k+1})\sigma_k\sigma_{k+1} = \alpha + \sigma_k\sigma_{k+1}$, we obtain,

$$\zeta_{nk}^g = \frac{1}{Q_n^g}\left[\alpha \sum_{\sigma_1 \cdots \sigma_k} \prod_{i=1}^{k-1}(1+\alpha\sigma_i\sigma_{i+1}) \prod_{j=1}^{k}(1+\gamma\sigma_j) \sum_{\sigma_{k+1}\cdots\sigma_n} \prod_{i=k+1}^{n-1}(1+\alpha\sigma_i\sigma_{i+1}) \prod_{j=k+1}^{n}(1+\gamma\sigma_j)\right.$$

$$+ \sum_{\sigma_1 \cdots \sigma_k} \prod_{i=1}^{k-1}(1+\alpha\sigma_i\sigma_{i+1}) \prod_{j=1}^{k}(1+\gamma\sigma_j) \sigma_k \sum_{\sigma_{k+1}\cdots\sigma_n} \prod_{i=k+1}^{n-1}(1+\alpha\sigma_i\sigma_{i+1}) \prod_{j=k+1}^{n}(1+\gamma\sigma_j) \sigma_{k+1}\left.\right]$$

That is,

$$\zeta_{nk}^g = \frac{1}{Q_n^g}\left(Y_k^g Y_{n-k}^g + \alpha Q_k^g Q_{n-k}^g\right) \quad (A7.1)$$

Obviously,

$$\zeta_{nk}^g = \zeta_{nn-k+1}^g \quad (A7.2)$$

# Appendix 8: Derivation and Solutions of the Homogeneous Linear Equations for $\delta_{nk}^g$

In this appendix, according to Eq.A6.7, we derive sufficient small $\delta_{nk}^g$ ($k = 1, \cdots n$) to evolve with $t$, which satisfy homogeneous linear equations, and give their general solutions.

## Appendix 8.1: Derivation and Solution of the Homogeneous Linear Equation for $\delta_1^{g1}$

Substituting $\eta_{1e}^g = s_{11}^{ge}$, $s_1^{g1} = s_{11}^{ge} + \delta_{11}^g$, $s_{11}^{ge} = \gamma_e$, $\gamma = \gamma_e + a_1^g \delta_{11}^g$, $a_1^g \equiv \dfrac{\partial \gamma_e}{\partial s_{11}^{ge}} = \dfrac{\partial \gamma_e}{\partial \theta_e} \dfrac{\partial \theta_e}{\partial \eta_{1e}^g} \dfrac{\partial \eta_{1e}^g}{\partial s_{11}^{ge}}$, $\dfrac{\partial \gamma_e}{\partial \theta_e} = 1 - \gamma_e^2$, $\dfrac{\partial \theta_e}{\partial \eta_{1e}^g} = \dfrac{A_1^g}{T}$, and $A_1^g \equiv \left(1 - \dfrac{1}{n}\right)\Theta_1^g = 0$ into Eq.A6.7A, we obtain,

$$\frac{1}{v}\frac{d\delta_{11}^g}{dt} = -\delta_{11}^g \tag{A8.1A}$$

The general solution of this linear homogeneous differential equation is,

$$\delta_{11}^g = \delta_{11}^g(0)\exp\left(-\frac{t}{\tau_{11}^g}\right) \tag{A8.1B}$$

where $\delta_{11}^g(0)$ is the value of $\delta_{11}^g$ at $t = 0$, and

$$\tau_{11}^g = \frac{1}{v} \tag{A8.1C}$$

## Appendix 8.2: Derivation and Solutions of the Homogeneous Linear Equations for $\delta_2^{gk}$

Substituting $s_{2k}^g = s_{2k}^{ge} + \delta_{2k}^g$; $\gamma = \gamma_e + a_2^g(\delta_2^{g1} + \delta_2^{g2})$, $a_2^g \equiv \dfrac{\partial \gamma_e}{\partial s_{2k}^{ge}} = \dfrac{\partial \gamma_e}{\partial \theta_e}\dfrac{\partial \theta_e}{\partial \eta_{2e}^g}\dfrac{\partial \eta_{2e}^g}{\partial s_{2k}^{ge}}$, $A_2^g \equiv \dfrac{1}{2}\Theta_2^g$, $\Theta_2^g = \dfrac{g}{2}\Theta_J$; $\zeta_{21}^g = \zeta_{21}^{ge} + b_{21}^g a_2^g(\delta_{21}^g + \delta_{22}^g)$, $b_{21}^g \equiv \dfrac{\partial \zeta_{21}^{ge}}{\partial \gamma_e}$, and $\zeta_{21}^{ge} = \zeta_{21}^g\big|_{\gamma=\gamma_e}$ into Eq.A6.7B, we get,

$$\frac{1}{v}\frac{d}{dt}\begin{bmatrix}\delta_{21}^g \\ \delta_{22}^g\end{bmatrix} = -[M]_{22}\begin{bmatrix}\delta_{21}^g \\ \delta_{22}^g\end{bmatrix} \tag{A8.2A}$$

where $[M]_{22} = \begin{bmatrix} 1 & -\alpha \\ -\alpha & 1 \end{bmatrix} - \begin{bmatrix} d_{21}^g & d_{21}^g \\ d_{22}^g & d_{22}^g \end{bmatrix}$ and $d_{21}^g = d_{22}^g \equiv a_2^g[1 - \alpha(\zeta_{21}^{ge} + \gamma_e b_{21}^g)]$.

The general solutions of the linear homogeneous differential equations are,

$$\begin{bmatrix} \delta_{21}^g \\ \delta_{22}^g \end{bmatrix} = e^{-\lambda_2^g vt} \begin{bmatrix} V_{21}^g \\ V_{22}^g \end{bmatrix} \tag{A8.2B}$$

Among them, $[V_2^{g1}, V_2^{g2}]$ is the eigenvector of $\lambda_2^g$.

Substituting this general solutions into Eq.A8.2A, it is obtained,

$$\det \begin{vmatrix} \lambda_2^g - 1 + d_{21}^g & \alpha + d_{21}^g \\ \alpha + d_{21}^g & \lambda_2^g - 1 + d_{21}^g \end{vmatrix} = 0 \tag{A8.2C}$$

and the two eigenvalues of $\lambda_2^g$ are,

$$\begin{cases} \lambda_{21}^g = 1 + \alpha \\ \lambda_{22}^g = 1 - \alpha - 2d_{21}^g \end{cases} \tag{A8.2D}$$

For $\lambda_{21}^g = 1 + \alpha$, the corresponding eigenvector $[V_{21}^{g1}, V_{21}^{g2}]$ satisfy the following equations,

$$\lambda_{21}^g \begin{bmatrix} V_{21}^{g1} \\ V_{21}^{g2} \end{bmatrix} = \begin{bmatrix} 1 - d_{21}^g & -\alpha - d_{21}^g \\ -\alpha - d_2^{g1} & 1 - d_{21}^g \end{bmatrix} \begin{bmatrix} V_{21}^{g1} \\ V_{21}^{g2} \end{bmatrix} \tag{A8.2E}$$

and,

$$\begin{bmatrix} V_{21}^{g1} \\ V_{21}^{g2} \end{bmatrix} = V_{22}^{g1} \begin{bmatrix} -1 \\ 1 \end{bmatrix} \tag{A8.2F}$$

For $\lambda_{22}^g = 1 - \alpha - 2d_{21}^g$, the corresponding eigenvector $[V_{22}^{g1}, V_{22}^{g2}]$ obey the following equations,

$$\lambda_{22}^g \begin{bmatrix} V_{22}^{g1} \\ V_{22}^{g2} \end{bmatrix} = \begin{bmatrix} 1 - d_{21}^g & -\alpha - d_{21}^g \\ -\alpha - d_{21}^g & 1 - d_{21}^g \end{bmatrix} \begin{bmatrix} V_{22}^{g1} \\ V_{22}^{g2} \end{bmatrix} \tag{A8.2G}$$

and,

$$\begin{bmatrix} V_{22}^{g1} \\ V_{22}^{g2} \end{bmatrix} = V_{22}^{g2} \begin{bmatrix} 1 \\ 1 \end{bmatrix} \tag{A8.2H}$$

where $[-1,1]$ and $[1,1]$ are respectively reduced eigenvectors of $\lambda_{21}^g$ and $\lambda_{22}^g$.

Therefore,

$$\begin{bmatrix} \delta_{21}^g \\ \delta_{22}^g \end{bmatrix} = V_{21}^{g2} \begin{bmatrix} -1 \\ 1 \end{bmatrix} \exp\left(\frac{t}{\tau_{21}^g}\right) + V_{22}^{g2} \begin{bmatrix} 1 \\ 1 \end{bmatrix} \exp\left(\frac{t}{\tau_{22}^g}\right) \tag{A8.2I}$$

and,

$$\tau_{2k}^g \equiv \frac{1}{\nu \lambda_{2k}^g}, k = 1,2 \tag{A8.2J}$$

## Appendix 8.3: Derivation and Solutions of the Homogeneous Linear Equations for $\delta_n^{gk}$

Substituting $s_{nk}^g = s_{nk}^{ge} + \delta_{nk}^g$; $\gamma = \gamma_e + a_n^g \sum_{k=1}^n \delta_{nk}^g$, $a_n^g \equiv \frac{\partial \gamma_e}{\partial s_{nk}^{ge}} = \frac{\partial \gamma_e}{\partial \theta_e} \frac{\partial \theta_e}{\partial \eta_{ne}^g} \frac{\partial \eta_{ne}^g}{\partial s_{nk}^{ge}}$,

$\frac{\partial \theta_e}{\partial \eta_{ne}^g} = \frac{A_n^g}{T}$, $\frac{\partial \eta_{ne}^g}{\partial s_{nk}^{ge}} = \frac{1}{n}$, $A_n^g \equiv \left(1 - \frac{1}{n}\right) \Theta_n^g$; $\zeta_{nk}^g = \zeta_{nk}^{ge} + b_{nk}^g a_n^g \sum_{k=1}^n \delta_{nk}^g$, $b_{nk}^g \equiv \frac{\partial \zeta_{nk}^{ge}}{\partial \gamma_e}$,

$\frac{\partial \zeta_{nk}^{ge}}{\partial s_{nk}^{ge}} = \frac{\partial \zeta_{nk}^{ge}}{\partial \gamma_e} \frac{\partial \gamma_e}{\partial s_{nk}^{ge}} = b_{nk}^g a_n^g$, and $\zeta_{nk}^{ge} = \zeta_{nk}^g \big|_{\gamma = \gamma_e}$ into Eq.A6.7C, we obtain,

$$\frac{1}{\nu} \frac{d}{dt} \begin{bmatrix} \delta_{n1}^g \\ \vdots \\ \delta_{nn}^g \end{bmatrix} = - \begin{bmatrix} M_{11} & \cdots & M_{1n} \\ \vdots & \ddots & \vdots \\ M_{n1} & \cdots & M_{nn} \end{bmatrix} \begin{bmatrix} \delta_{n1}^g \\ \vdots \\ \delta_{nn}^g \end{bmatrix} \tag{A8.3A}$$

and the square matrix is,

$$[M]_{nn} = -\begin{bmatrix} -1 & \alpha & 0 & & & \\ \beta & -1 & \beta & \cdots & & 0 \\ 0 & \beta & -1 & & & \\ \vdots & & & \ddots & & \vdots \\ & & & -1 & \beta & 0 \\ 0 & & \cdots & \beta & -1 & \beta \\ & & & 0 & \alpha & -1 \end{bmatrix} - \begin{bmatrix} d_{n1}^g & d_{n1}^g & & d_{n1}^g & d_{n1}^g \\ d_{n2}^g & d_{n2}^g & & d_{n2}^g & d_{n2}^g \\ \vdots & & \ddots & & \vdots \\ d_{nn-1}^g & d_{nn-1}^g & \cdots & d_{nn-1}^g & d_{nn-1}^g \\ d_{nn}^g & d_{nn}^g & & d_{nn}^g & d_{nn}^g \end{bmatrix}$$

$$\begin{cases} d_{n1}^g \equiv a_n^g[1 - \alpha(\zeta_{n1}^{ge} + \gamma_e b_{n1}^g)] \\ d_{nn}^g \equiv a_n^g[1 - \alpha(\zeta_{nn-1}^{ge} + \gamma_e b_{nn-1}^g)] \\ d_{nk}^g \equiv a_n^g[1 - \beta(\zeta_{nk-1}^{ge} + \zeta_{nk+1}^{ge}) - \beta\gamma_e(b_{nk-1}^g + b_{nk}^g)] \end{cases} \tag{A8.3B}$$

where $k = 2, \cdots n - 1$.

The general solutions of this linear homogeneous differential equations are,

$$\begin{bmatrix} \delta_{n1}^g \\ \vdots \\ \delta_{nn}^g \end{bmatrix} = e^{-\lambda_n^g \nu t} \begin{bmatrix} V_n^{g1} \\ \vdots \\ V_n^{gn} \end{bmatrix} \tag{A8.3C}$$

Substituting this general solution into Eq.A8.3A, we obtain,

$$\lambda_n^g \begin{bmatrix} V_n^{g1} \\ \vdots \\ V_n^{gn} \end{bmatrix} = \begin{bmatrix} M_{11} & \cdots & M_{1n} \\ \vdots & \ddots & \vdots \\ M_{n1} & \cdots & M_{nn} \end{bmatrix} \begin{bmatrix} V_n^{g1} \\ \vdots \\ V_n^{gn} \end{bmatrix} \tag{A8.3D}$$

The conditions that the homogeneous linear equations have non-zero eigenvectors are,

$$\det|[M]_{nn} - \lambda_n^g [I]_{nn}| = 0 \tag{A8.3E}$$

and this equation is called the characteristic equation that $\lambda_n^g$ satisfies, where $[I]_{nn}$ is an n-order unit square matrix.

The $n$ eigenvalues $(\lambda_{n1}^g, \cdots \lambda_{nn}^g)$ of $[M]_{nn}$ can be obtained from Eq.A8.3E, and the eigenvector $[V_{nk}^{g1}, \cdots V_{nk}^{gn}]$ corresponding to $\lambda_{nk}^g$ satisfies the following equation,

$$\lambda_{nk}^g \begin{bmatrix} V_{nk}^{g1} \\ \vdots \\ V_{nk}^{gn} \end{bmatrix} = \begin{bmatrix} M_{11} & \cdots & M_{1n} \\ \vdots & \ddots & \vdots \\ M_{n1} & \cdots & M_{nn} \end{bmatrix} \begin{bmatrix} V_{nk}^{g1} \\ \vdots \\ V_{nk}^{gn} \end{bmatrix} \tag{A8.3F}$$

Since Eq.A8.3F is a homogeneous linear equation, only $n-1$ values in $[V_{nk}^{g1}, \cdots V_{nk}^{gn}]$ can be determined. Defining $v_{nk}^{gi} \equiv \frac{V_{nk}^{gi}}{V_{nk}^{gn}}, i = 1, \cdots n$, the reduced eigenvector $[v_{nk}^{g1}, \cdots 1]$ of $\lambda_{nk}^g$ satisfies the following equation,

$$\begin{bmatrix} M_{11} - \lambda_{nk}^g & \cdots & M_{1n-1} \\ \vdots & \ddots & \vdots \\ M_{n-11} & \cdots & M_{n-1n-1} - \lambda_{nk}^g \end{bmatrix} \begin{bmatrix} v_{nk}^{g1} \\ \vdots \\ v_{nk}^{gn-1} \end{bmatrix} = -\begin{bmatrix} M_{1n} \\ \vdots \\ M_{n-1n} \end{bmatrix} \tag{A8.3G}$$

Therefore,

$$\begin{bmatrix} \delta_{n1}^g \\ \vdots \\ \delta_{nn}^g \end{bmatrix} = \sum_{k=1}^n V_{nk}^{gn} \begin{bmatrix} v_{nk}^{g1} \\ \vdots \\ 1 \end{bmatrix} \exp\left(\frac{t}{\tau_{nk}^g}\right) \tag{A8.3H}$$

and,

$$\tau_{nk}^g \equiv \frac{1}{v\lambda_{nk}^g}, k = 1, \cdots n \tag{A8.3I}$$

## Appendix 9: Calculating Complex Susceptibility of n-g-PSSs

After the following stepless external electric field ($E$) small enough with $t$,

$$E = \begin{cases} E_0, & t \leq 0 \\ 0, & t > 0 \end{cases} \tag{A9.1}$$

is applied to 3D-RSIGM along the PS direction, the Hamiltonian of n-g-PSSs with $E$ and the corresponding partition function ($Z_{nE}^g$) are same as Eqs.A5.2A-B.

By expressing the values of $s_{nk}^g$, $\eta_n^g$, and $\theta$ at $t = 0$ as $s_{nk}^g(0)$, $\eta_n^g(0)$, and $\theta(0)$ respectively, we get,

$$s_{nk}^g(0) = \frac{1}{Z_n^g} \sum_{\sigma_1 \cdots \sigma_n} \sigma_k \exp\left[\varpi \sum_{i=1}^{n-1} \sigma_i \sigma_{i+1} + \theta^E \sum_{j=1}^{n} \sigma_j\right] + O(E^2) \tag{A9.2}$$

Let the deviation of $s_{nk}^g(0)$ from $s_{nk}^{ge}$ induced by $E_0$ be $\delta_{nk}^g(0)$, i.e. $\delta_{nk}^g(0) \equiv s_{nk}^g(0) - s_{nk}^{ge}$, then,

$$\delta_{nk}^g(0) = \frac{\partial s_{nk}^g(0)}{\partial E_0} E_0 = \frac{\partial s_{nk}^g(0)}{\partial \theta^E(0)} \left(\frac{\partial \theta(0)}{\partial E_0} + \frac{\mu}{k_B T}\right) E_0$$

$$= \frac{\partial s_{nk}^{ge}}{\partial \theta_e} \left(\frac{\mu}{\varepsilon_0} \frac{\partial \eta_n^g(0)}{\partial E_0} \frac{A_n^g}{T} + \frac{C_w}{N_0 T}\right) \frac{\varepsilon_0 E_0}{\mu}$$

By $\left.\frac{\partial s_{nk}^g(0)}{\partial \theta^E(0)}\right|_{E_0 \to 0} = \frac{\partial s_{nk}^{ge}}{\partial \theta_e} = \frac{\partial s_{nk}^{ge}}{\partial \gamma_e} \frac{\partial \gamma_e}{\partial \theta_e} = (1 - \gamma_e^2) \frac{\partial s_{nk}^{ge}}{\partial \gamma_e}$, we obtain,

$$\delta_{nk}^g(0) = (1 - \gamma_e^2)\left(\frac{A_n^g \chi_s^{ng}}{T} + \frac{C_w}{N_0 T}\right) \frac{\partial s_{nk}^{ge}}{\partial \gamma_e} \frac{\varepsilon_0 E_0}{\mu} \tag{A9.3}$$

Substitute this equation into Eq.A8.3H, we get the following equations that $V_{nk}^{gn}$ ($k = 1, \cdots n$) obey,

$$\sum_{k=1}^{n} \begin{bmatrix} v_{nk}^{g1} \\ \vdots \\ 1 \end{bmatrix} \frac{\mu V_{nk}^{gn}}{\varepsilon_0 E_0} = (1 - \gamma_e^2)\left(\frac{A_n^g \chi_s^{ng}}{T n} + \frac{C_w}{N_0 T}\right) \begin{bmatrix} \partial s_{n1}^{ge}/\partial \gamma_e \\ \vdots \\ \partial s_{nn}^{ge}/\partial \gamma_e \end{bmatrix} \tag{A9.4}$$

Therefore, for $t \geq 0$, the polarization ($P_n^g$) of n-g-PSSs with $t$ is,

$$P_n^g = \mu \sum_{i=1}^{n} \delta_{ni}^g = \varepsilon_0 E_0 \sum_{k=1}^{n} \Delta_{nk}^g e^{-t/\tau_{nk}^g} \tag{A9.5}$$

and,

$$\Delta_{nk}^g \equiv \frac{V_{nk}^{gn}}{\varepsilon_0 E_0} \sum_{i=1}^{n} v_{nk}^{gi} \tag{A9.6}$$

By Eqs.A9.5-6, it is obtained that the linear complex susceptibility ($\chi_n^{g*}$) of n-g-PSSs,

$$\chi_n^{g*} = \chi_n^{g\prime} - \chi_n^{g\prime\prime} = \sum_{k=1}^{n} \frac{\Delta_{nk}^g}{1 + i_c \omega \tau_{nk}^g} \tag{A9.7}$$

Obviously,

$$\chi_s^{ng} = \sum_{k=1}^{n} \Delta_{nk}^g \tag{A9.8}$$

# Appendix 10: Burns Transformation of High Temperature Thermal Strain in PRFEs

At present, the experimental results of the Burns transformation of the high-temperature thermal strain ($s_{kl}^T$, $k,l = 1,2,3$) in PRFEs are explained based on the macroscopic quadradic-electrostrictive (QES) effect, that is, the crystal strains ($s_{kl}^{QES}$, $k,l = 1,2,3$) induced by the spontaneous polarization component ($P_s^i$, $i = 1,2,3$) are [5],

$$s_{kl}^{QES} = \sum_{i,j=1}^{3} q_{kl}^{ij} P_s^i P_s^j \tag{A10.1}$$

where $q_{kl}^{ij}$ is the quadradic electrostrictive coefficient.

So, the thermal strains ($s_{kl}^T$) of PRFEs are,

$$s_{kl}^T = s_{kl}^0 + s_{kl}^{QES} \tag{A10.2}$$

and $s_{kl}^0$ are the thermal strains caused by the non-harmonic portion of the interaction that constructs the crystal lattice.

When the temperature is high, $s_{k,l}^0$ satisfies the linear relationship with $T$, and we get,

$$s_{kl}^T - s_{kl}^0(T_r) = \alpha_{kl}(T - T_r) + s_{kl}^{QES} \tag{A10.3}$$

Among them, $T_r$ is a reference temperature, $\alpha_{kl}$ is the high temperature coefficient of thermal expansion caused by the non-harmonic interaction, and it is nearly independent of $T$.

Based on Eqs.A10.1-3, Cross et al. [5] think that the deviation of $s_{kl}^T$ from the high-temperature linear behavior in RFFEs is due to the emergence of PNRs first proposed by Burns et al. [80-82], which is much higher than that of RFPT,

It is conceivable that the variation in the local-interaction (LI) energy of the nearest neighbor PSs inevitably induces a change in the relative position of the ions in the unit cell, that is, the local distortion (LD) of the crystal lattice, which is called as the LI-LD coupling here.

It can also be expected that the LD will inevitably lead to the change of the interaction energy constant ($J$) between PSs. In this paper, we refer the LI-LD coupling is respectively the strong or weak coupling depending on that it changes or does not change the crystal lattice symmetry. Under the weak and linear LI-LD coupling approximation, the local distortion ($s_{k,l}^{i,j}$) caused by the interaction between $i^{th}$ and nearest neighbor $j^{th}$ PS is,

$$s_{kl}^{ij} = c_{kl}\sigma_i\sigma_j r_i^{(\phi)} r_j^{(\phi)} \tag{A10.4}$$

where $c_{kl}$ is the coefficient of LI-LD coupling.

Then, the strain ($s_{kl}^{LI-LD}$) of PRFEs caused by the LI-LD coupling is,

$$s_{kl}^{LC-LD} \approx \frac{1}{N}\sum_{ij}^{\{nn\}} s_{kl}^{ij} = -c_{kl}\frac{u_{ps}}{J} \tag{A10.5}$$

Therefore, the high temperature $s_{kl}^T$ of PRFEs is,

$$s_{kl}^T - s_{kl}^0(T_r) \approx \alpha_{kl}(T-T_r) - c_{kl}\frac{u_{ps}}{J} \tag{A10.6}$$

It is worth pointing out that, based on the Weiss mean-field of three-dimensional Ising model, $u_{ps} = -3J\left(\frac{P_s}{N_0\mu}\right)^2$, and it can be obtained from Eq.A10.6,

$$s_{kl}^T - s_{kl}^0(T_r) \approx \alpha_{kl}(T-T_r) + 3c_{kl}J\left(\frac{P_s}{N_0\mu}\right)^2 \tag{A10.7}$$

Which is in agreement with Eq.A10.3. This seems to suggest that the explanation of the Burns transformation of thermal strain in PRFEs based on the macroscopic quadratic electrostrictive effect may include larger errors.

## Appendix 11: Burns Transformation of High Temperature Refractive Index in PRFEs

Currently, the Burns transformation of the high-temperature refractive index ($n_{kl}$, $k,l = 1,2,3$) in PRFEs are explained by the macroscopic Kerr (quadradic electrooptic) effect, that is, the refractive indexes ($n_{kl}^{KE}$, $k,l = 1,2,3$) induced by $P_s^i$ ($i = 1,2,3$) are [80-82],

$$n_{kl}^{KE} = \sum_{i,j=1}^{3} K_{kl}^{ij} P_s^i P_s^j \tag{A11.1}$$

where $K_{kl}^{ij}$ is the Kerr coefficient.

So, $n_{kl}$ ($k,l = 1,2,3$) of PRFEs is,

$$n_{kl} = n_{kl}^0 + n_{kl}^{KE} \tag{A11.2}$$

and $n_{kl}^0$ is the refraction index that is not related to the correlation between PSs.

At high temperature, $n_{kl}^0$ meets the linear relationship with temperature and,

$$n_{kl} - n_{kl}^0(T_r) = b_{kl}(T - T_r) + n_{kl}^{KE} \tag{A11.3}$$

Among them, $b_{kl}$ is the thermo-optic coefficient that is caused by the non-harmonic interaction.

Based on Eqs.A11.1-3, Burns et al. [80-82] considered for the first time that the deviation of $n_{kl}$ to the high temperature linear behavior of PRFEs is caused by the emergence of PNRs on cooling, which is much higher than that of RFPT.

It is conceivable that the local distortion of the crystal lattice caused by the LI-LD coupling (Appendix 10) will inevitably lead to the change of the local-electron-clouds (LE) in the lattice, which is abbreviated as LI-LE coupling. The LI-LE coupling will also induce the change in the local optical frequency permittivity of PRFEs, and thus in the local refractive index.

It can also be expected that the interaction energy ($J$) between the PSs will be modified by the changes of the local electron clouds. In this paper, it is called the strong or weak LI-LE coupling that change or do not change the symmetry of the crystal lattice, respectively.

Under the weak and linear LI-LE coupling approximation, the induced local refractive index ($n_{kl}^{ij}$) by the interaction between the i$^{th}$ and the nearest-neighbor j$^{th}$ PS is,

$$n_{kl}^{ij} = d_{kl}\sigma_i\sigma_j r_i^{(\phi)} r_j^{(\phi)} \tag{A11.4}$$

where $d_{kl}$ is the LI-LE coupling coefficient.

The refractive index ($n_{kl}^{LI-LC}$) of PRFEs due to LI-LE coupling is,

$$n_{kl}^{LI-LC} \approx \frac{1}{N}\sum_{i,j}^{\{nn\}} n_{kl}^{ij} = -d_{kl}\frac{u_{ps}}{J} \tag{A11.5}$$

Therefore, the high temperature $n_{kl}$ is,

$$n_{kl} - n_{kl}^0(T_r) \approx b_{kl}(T - T_r) - d_{kl}\frac{u_{ps}}{J} \tag{A11.6}$$

According to the internal energy results of Weiss mean-field in three-dimensional Ising model (Appendix 10) and Eq.A11.6, we obtain,

$$n_{kl} - n_{kl}^0(T_r) \approx b_{kl}(T - T_r) + 3d_{kl}J\left(\frac{P_s}{N_0\mu}\right)^2 \tag{A11.7}$$

which is consistent with Eq.A11.3. This seems to indicate that, the interpretation of the Burns transition of the refractive index in PRFEs based on the macroscopic Kerr effect may include large errors.

## References

[1] G. A. Smolenskii, and K. I. Rozgachev, Zhurnal Tekhnicheskoi Fiziki **24**, 1751 (1954).
[2] G. A. Smolenskii *et al.*, Soviet Physics-Solid State **2**, 2651 (1961).
[3] G. A. Smolenskii, J. Phys. Soc. Jpn **28**, 26 (1970).
[4] X. Yao, Z. L. Chen, and L. E. Cross, J. Appl. Phys. **54**, 3399 (1983).

[5] L. E. Cross, Ferroelectrics **76**, 241 (1987).

[6] A. A. Bokov, and Z. G. Ye, J. Mater. Sci. **41**, 31 (2006).

[7] Z. Kutnjak, J. Petzelt, and R. Blinc, Nature **441**, 956 (2006).

[8] S. Tinte *et al.*, Phys. Rev Lett. **97**, 137601 (2006).

[9] M. McQuarrie and F. W. Behnke, J. Am. Ceram. Soc. **37**, 539 (1954).

[10] I. Grinberg, Y. Shin, and A. M. Rappe, Phys. Rev Lett. **103**, 197601 (2009).

[11] R. A. Cowley *et al.*, Adv. Phys. **60**, 229 (2011).

[12] V. V. Shvartsman, and D. C. Lupascu, J. Am. Ceram. Soc. **95**, 1 (2012).

[13] N. Novak *et al.*, Phys. Rev Lett. **109**, 037601 (2012).

[14] A. R. Akbarzadeh *et al.*, Phys. Rev Lett. **108**, 257601 (2012).

[15] H. Takenaka, I. Grinberg, and A. M. Rappe, Phys. Rev Lett. **110**, 147602 (2013).

[16] G. Geneste, L. Bellaiche, and J. Kiat, Phys. Rev Lett. **116**, 247601 (2016).

[17] V. Westphal, W. Kleemann, and M. D. Glinchuk, Phys. Rev Lett. **68**, 847 (1992).

[18] W. Kleemann, and A. Kloessner, Ferroelectrics **150**, 35 (1993).

[19] M. D. Glinchuk, and R. Farhi, J. Phys-Condens. Mat. **8**, 6985 (1996).

[20] M. D. Glinchuk, British Ceramic Transactions **103**, 76 (2004).

[21] D. Viehland *et al.*, J. Appl. Phys. **68**, 2916 (1990).

[22] D. Viehland, M. Wuttig, and L. E. Cross, Ferroelectrics **120**, 71 (1991).

[23] D. Viehland *et al.*, Phys. Rev. B **43**, 8316 (1991).

[24] R. Pirc, and R. Blinc, Phys. Rev. B **60**, 13470 (1999).

[25] R. Blinc *et al.*, Phys. Rev Lett. **83**, 424 (1999).

[26] N. Setter and L. E. Cross, J. Appl. Phys. **51**, 4356 (1980).

[27] N. Setter, and L. E. Cross, J. Mater. Sci. **15**, 2478 (1980).

[28] N. Demathan *et al.*, J. Phys-Condens. Mat. **3**, 8159 (1991).

[29] E. V. Colla *et al.*, J. Phys-Condens. Mat. **4**, 3671 (1992).

[30] V. Bovtun *et al.*, Ferroelectrics **298**, 23 (2004).

[31] V. Bovtun *et al.*, J. Eur. Ceram. Soc. **26**, 2867 (2006).

[32] J. Banys *et al.*, Physica Status Solidi C **6**, 2725 (2009).

[33] E. Palaimiene *et al.*, J. Appl. Phys. **116**, 104103 (2014).

[34] D. Jablonskas *et al.*, Appl. Phys. Lett. **107**, 142905 (2015).

[35] W. H. Huang, D. Viehland, and R. R. Neurgaonkar, J. Appl. Phys. **76**, 490 (1994).

[36] J. Dec *et al.*, Eur. Phys. J B **14**, 627 (2000).

[37] T. Maiti, R. Guo, and A. S. Bhalla, J. Am. Ceram. Soc. **91**, 1769 (2008).

[38] D. Nuzhnyy *et al.*, Phys. Rev. B **86**, 014106 (2012).

[39] W. Kleemann *et al.*, Appl. Phys. Lett. **102**, 232907 (2013).

[40] R. Blinc, and B. Zeks, Adv. Phys. **21**, 693 (1972).

[41] W. Cochran, Phys. Rev Lett. **3**, 412 (1959).

[42] W. Cochran, Adv. Phys. **9**, 387 (1960).

[43] W. Cochran, Adv. Phys. **10**, 401 (1961).

[44] W. Cochran, Rep. Prog. Phys. **26**, 1 (1963).

[45] Z. Yu, R. Y. Guo, and A. S. Bhalla, J. Appl. Phys. **88**, 410 (2000).

[46] Z. Yu *et al.*, J. Appl. Phys. **92**, 1489 (2002).

[47] A. R. Akbarzadeh *et al.*, Phys. Rev. B **72**, 205104 (2005).

[48] W. Kleemann *et al.*, Appl. Phys. Lett. **104**, 182910 (2014).

[49] E. Cordfunke, and R. Konings, Thermochim. Acta **156**, 45 (1989).

[50] A. A. Dealcantara, A. Desouza, and F. Moreira, Phys. Rev. B **49**, 9206 (1994).

[51] B. Kutlu, and A. E. Genc, Physica A **392**, 451 (2013).

[52] R. J. Glauber, J. Math. Phys. **4**, 294 (1963).

[53] M. Ertas, B. Deviren, and M. Keskin, Phys. Rev. E **86**, 051110 (2012).

[54] P. V. Prudnikov *et al.*, Prog. Theor. Exp. Phys, 053A01 (2015).

[55] W. P. Mason, Phys. Rev. **72**, 854 (1947).

[56] E. Nakamura, and M. Hosoya, J. Phys. Soc. Jpn. **23**, 844 (1967).

[57] M. A. Girtu *et al.*, J. Appl. Phys. **81**, 4410 (1997).

[58] A. A. Belik *et al.*, Chem. Mater. **19**, 1679 (2007).

[59] J. A. Quilliam *et al.*, Phys. Rev Lett. **98**, 037203 (2007).

[60] N. Marcano *et al.*, Phys. Rev. B **76**, 224419 (2007).

[61] A. A. Belik, and E. Takayama-Muromachi, J. Phys-Condens. Mat. **20**, 25211 (2008).

[62] D. X. Li *et al.*, J. Appl. Phys. **103**, 7B (2008).

[63] A. Desouza, and F. Moreira, Europhys. Lett. **17**, 491 (1992).

[64] J. A. Plascak, W. Figueiredo, and B. Grandi, Braz. J. Phys. **29**, 579 (1999).

[65] E. Ising, Zeitschrift fur Physik **31**, 253 (1925).

[66] H. A. Kramers, and G. H. Wannier, Phys. Rev. **60**, 263 (1941).

[67] G. H. Wannier, Rev. Mod. Phys. **17**, 50 (1945).

[68] L. Onsager, Phys. Rev. **65**, 117 (1944).

[69] A. E. Ferdinand and M. E. Fisher, Phys. Rev. **185**, 832 (1969).

[70] G. F. Newell, and E. W. Montroll, Rev. Mod. Phys. **25**, 353 (1953).

[71] Y. IMRY, and S. MA, Phys. Rev Lett. **35**, 1399 (1975).

[72] J. Wang *et al.*, Phys. Rev. E **87**, 052107 (2013).

[73] K. Binder, W. Kinzel, and D. Stauffer, Zeitschrift fur Physik B-Condensed Matter **36**, 161 (1979).

[74] M. A. Neto, R. A. Dos Anjos, and J. Ricardo De Sousa, Phys. Rev. B **73**, 214439 (2006).

[75] P. Weiss, Physikalische Zeitschrift **9**, 358 (1908).

[76] Y. N. Huang, Y. N. Wang, and H. M. Shen, Phys. Rev. B **46**, 3290 (1992).

[77] Y. N. Huang *et al.*, Phys. Rev. B **55**, 16159 (1997).

[78] K. W. Wagner, Ann. Phys-Berlin **40**, 817 (1913).

[79] P. Debye, Ann. Phys-Berlin **33**, 441 (1910).

[80] G. Burns, and B. A. Scott, Solid State Commun. **13**, 423 (1973).

[81] G. Burns, and F. H. Dacol, Phys. Rev. B **28**, 2527 (1983).

[82] G. Burns, and F. H. Dacol, Ferroelectrics **104**, 25 (1990).

[83] P. Lehnen *et al.*, Eur. Phys. J. B **14**, 633 (2000).

[84] P. M. Gehring *et al.*, Phys. Rev. B **79**, 224109 (2009).

[85] C. Stock *et al.*, Phys. Rev. B **81**, 144127 (2010).

[86] R. Sommer, N. K. Yushin, and J. J. Vanderklink, Phys. Rev. B **48**, 13230 (1993).

[87] T. Granzow *et al.*, Phys. Rev Lett. **92**, 065701 (2004).

[88] X. Zhao *et al.*, Phys. Rev. B **75**, 104106 (2007).

[89] S. Miga *et al.*, Phys. Rev. B **80**, 220103 (2009).

[90] V. V. Shvartsman, B. Dkhil, and A. L. Kholkin, Annu. Rev. Mater. Res. **43**, 423 (2013).

[91] Y. Moriya *et al.*, Phys. Rev Lett. **90**, 205901 (2003).

[92] M. V. Gorev *et al.*, Phys. Solid State+ **47**, 2304 (2005).

[93] W. Kleemann *et al.*, Phys. Rev Lett. **97**, 65702 (2006).

[94] M. Tachibana *et al.*, Phys. Rev. B **80**, 94115 (2009).

Figures and Figure Captions

Fig.1

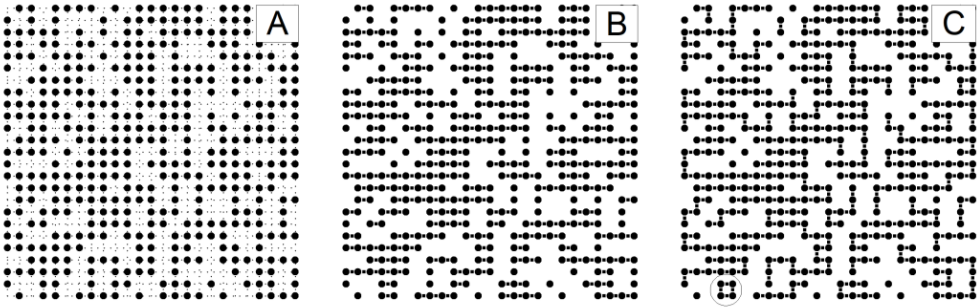

fig.1 A) Schematic distribution of PSs in a y-z-plane in 3D-RSIM of $\phi = 0.4$; B) Construction process of short PSSs along the z-axis direction of the crystal lattice; C) Connection process of nearest neighbor endpoints of the short PSSs to form long PSSs along the y-axis direction (An endpoint already connected to a PSS is no longer reconnected). It is a ring PSS in the circle of Fig.1C.

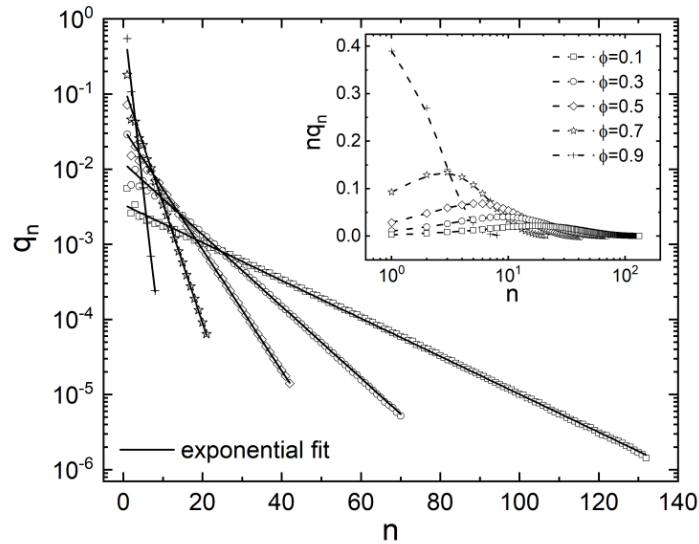

Fig.2 $q_n$ vs $n$ in 3D-RSIM with series of $\phi$. The inset shows $nq_n$ vs $n$ for the corresponding $\phi$ values.

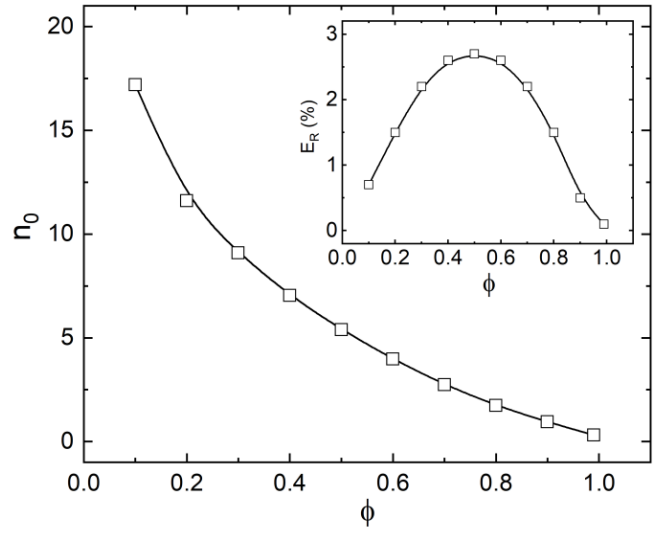

Fig.3 $n_0$ vs $\phi$ in 3D-RSIM. The inset shows $E_R$ vs $\phi$.

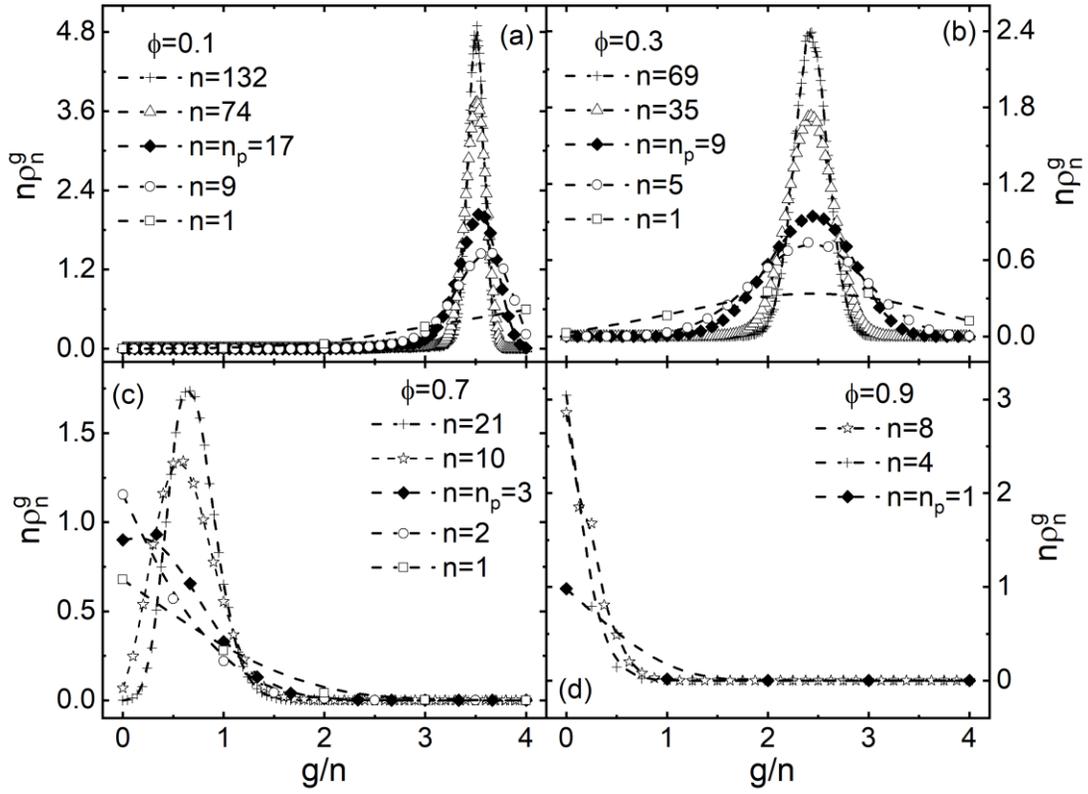

Fig.4 $n\rho_n^g$ vs $g/n$ and $n$ in 3D-RSIM with series of $\phi$.

Fig.5

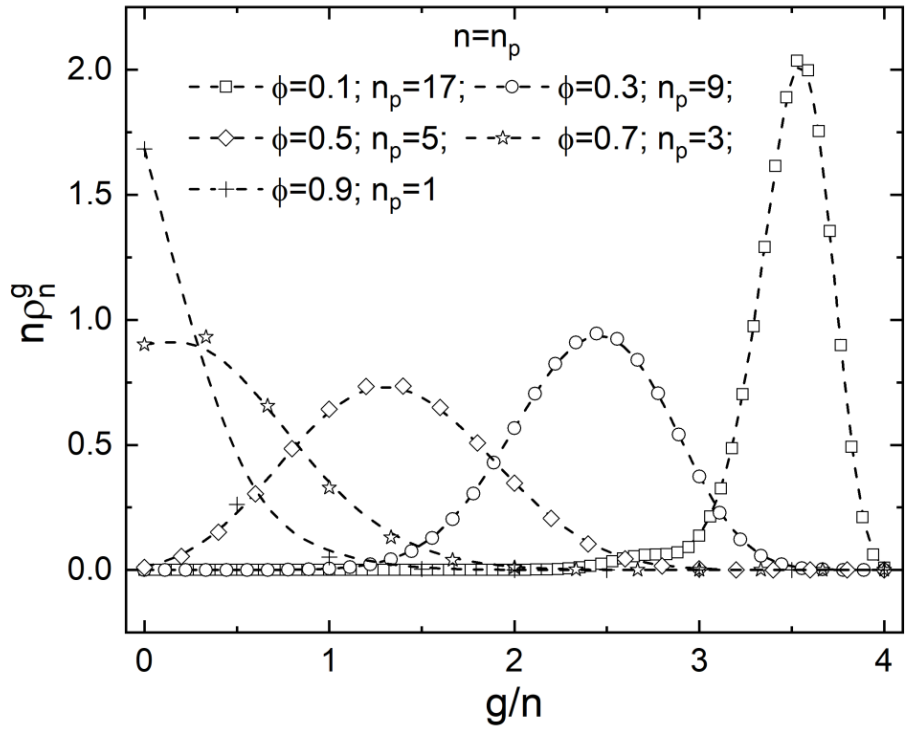

Fig.5 $n\rho_n^g$ of $n = n_p$ changes with $g/n$ in 3D-RSIM with series of ϕ.

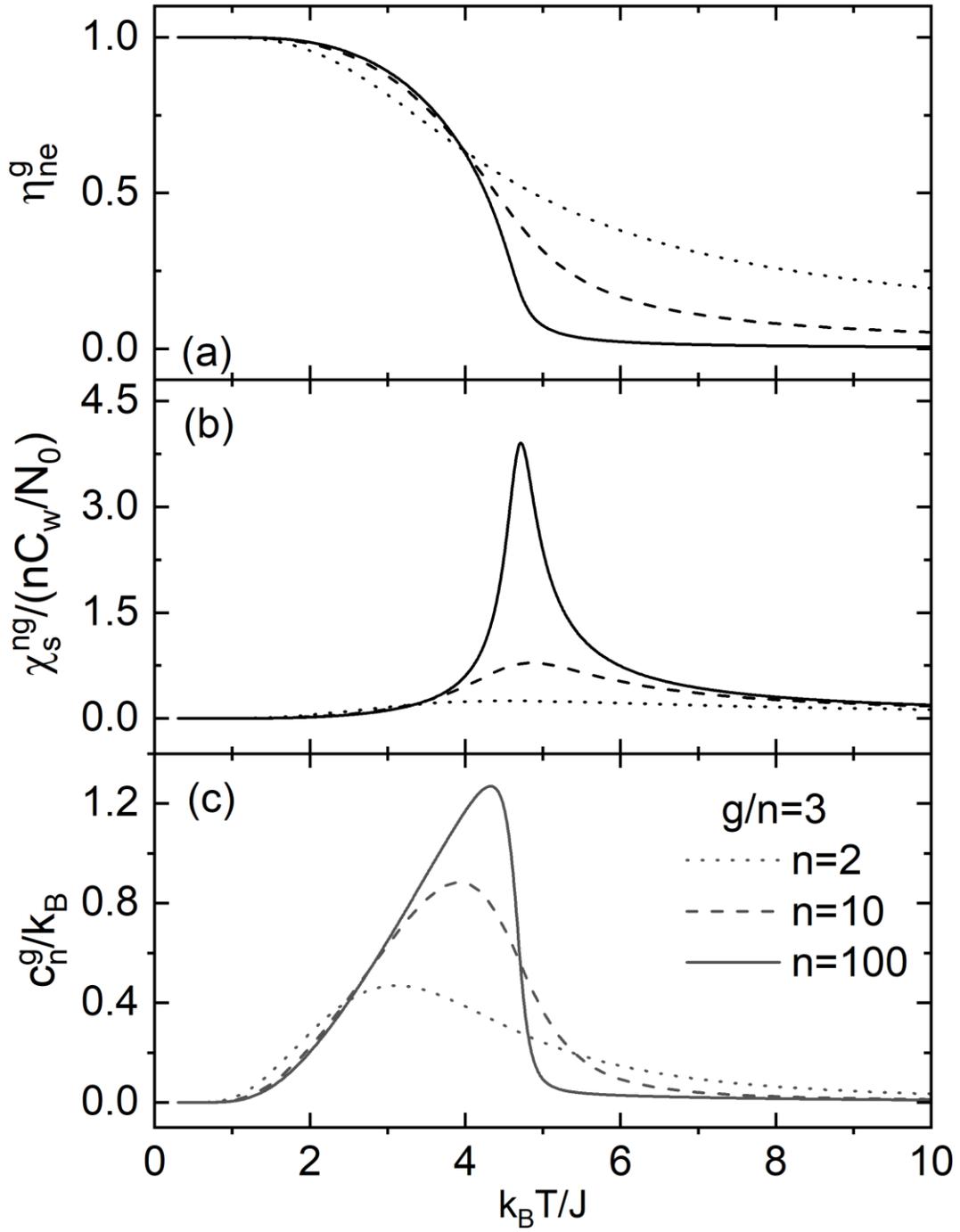

Fig.6 $\eta_{ne}^{g}$, $\chi_{s}^{ng}$, and $c_{n}^{g}$ of n-g-PSSs vs $T$ for $\frac{g}{n} = 3$ and series of $n$.

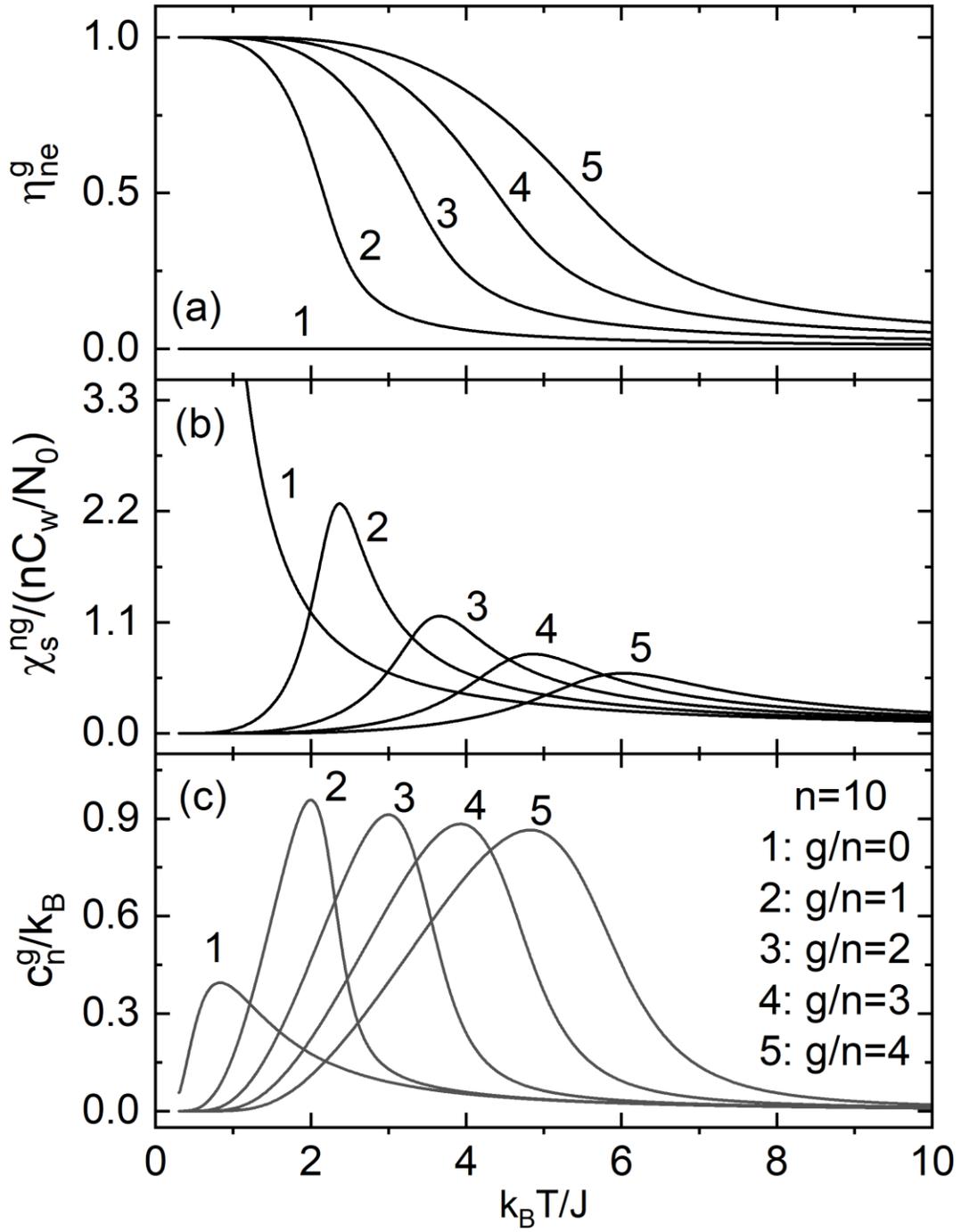

Fig.7 $\eta_{ne}^{g}$, $\chi_{s}^{ng}$, and $c_{n}^{g}$ of n-g-PSSs vs $T$ for $n = 10$ and series of $g$.

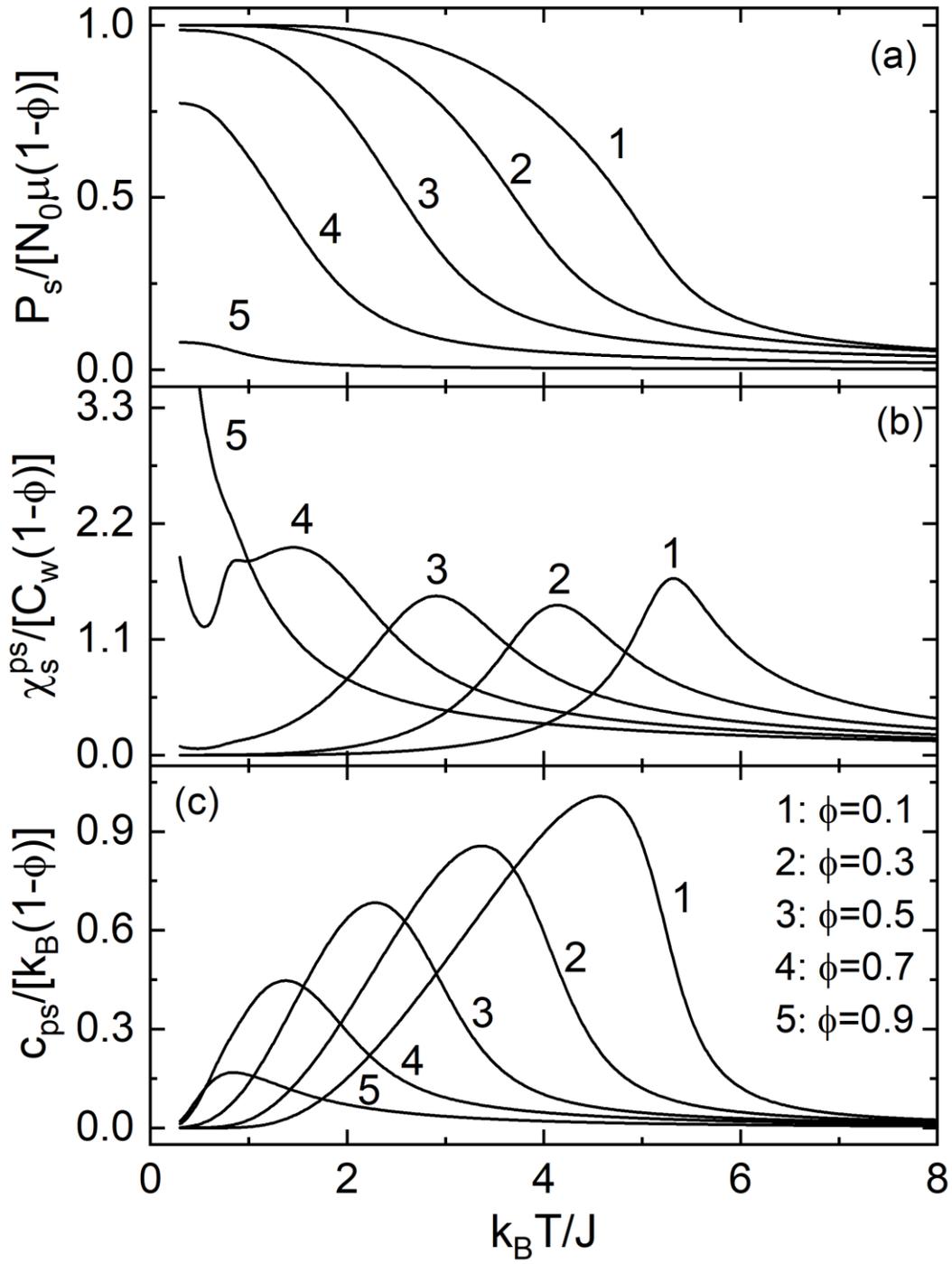

Fig.8 $P_s$, $\chi_s^{ps}$, and $c_{ps}$ of 3D-RSIM vs $T$ for series of ϕ.

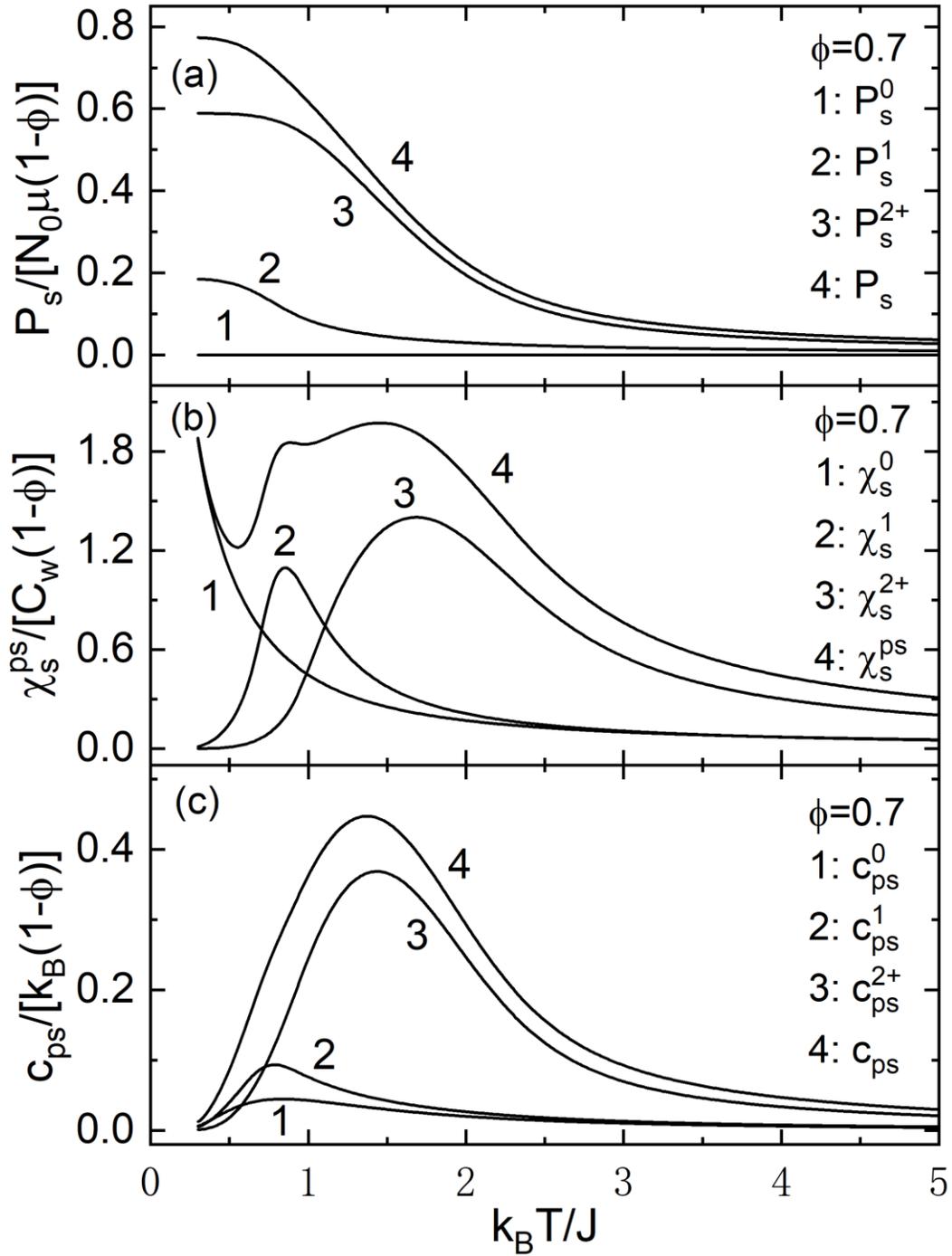

Fig.9 Spontaneous polarization, static susceptibility, and average specific heat per PS vs $T$ of 3D-RSIM as well as n-0-PSSs, n-1-PSSs, and n-$2^+$-PSSs in the model for $\phi = 0.7$.

Fig.10

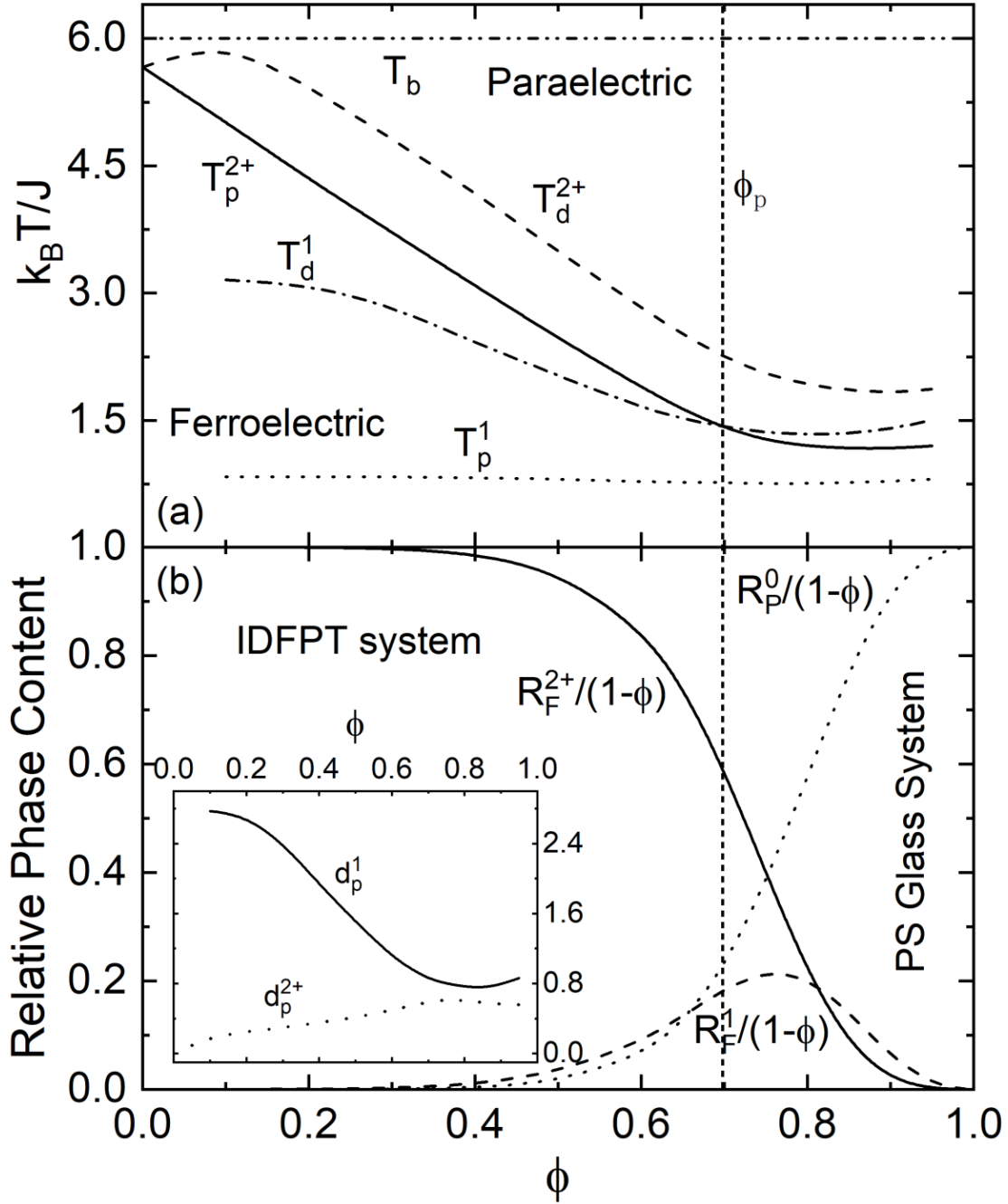

Fig.10 Phase diagram of 3D-RSIM, namely $T_p^1$, $T_p^{2+}$, $T_d^1$, $T_d^{2+}$, $d_p^1$, $d_p^{2+}$, $R_P^0$, $R_F^1$, $R_F^{2+}$, and $T_b$ vs $\phi$.

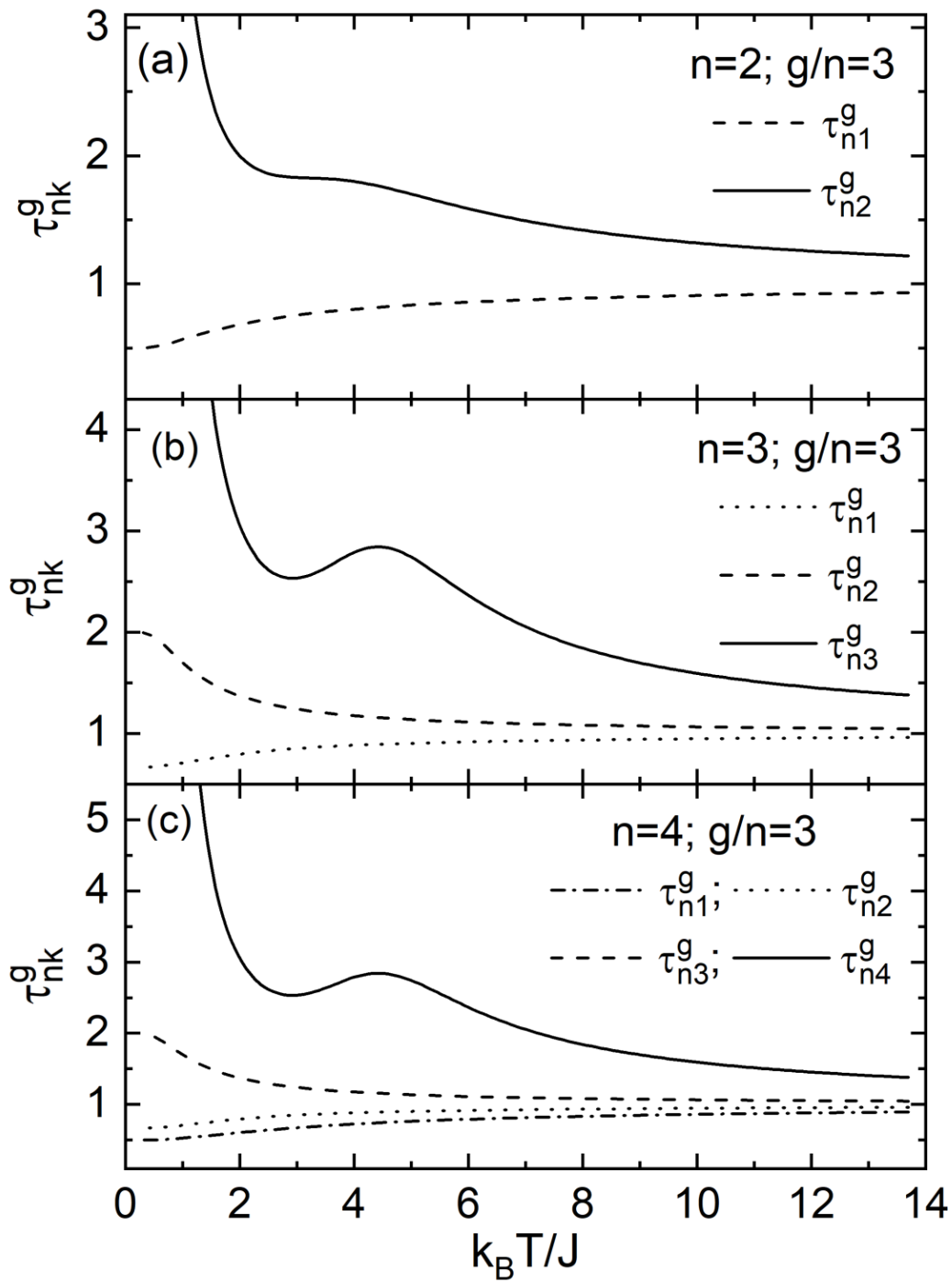

Fig.11 $\nu \tau_{nk}^{g}$ $(k = 1, \cdots n)$ of n-g-PSSs vs $T$ for $n = 2, 3, 4$ and $\frac{g}{n} = 3$.

Fig.12

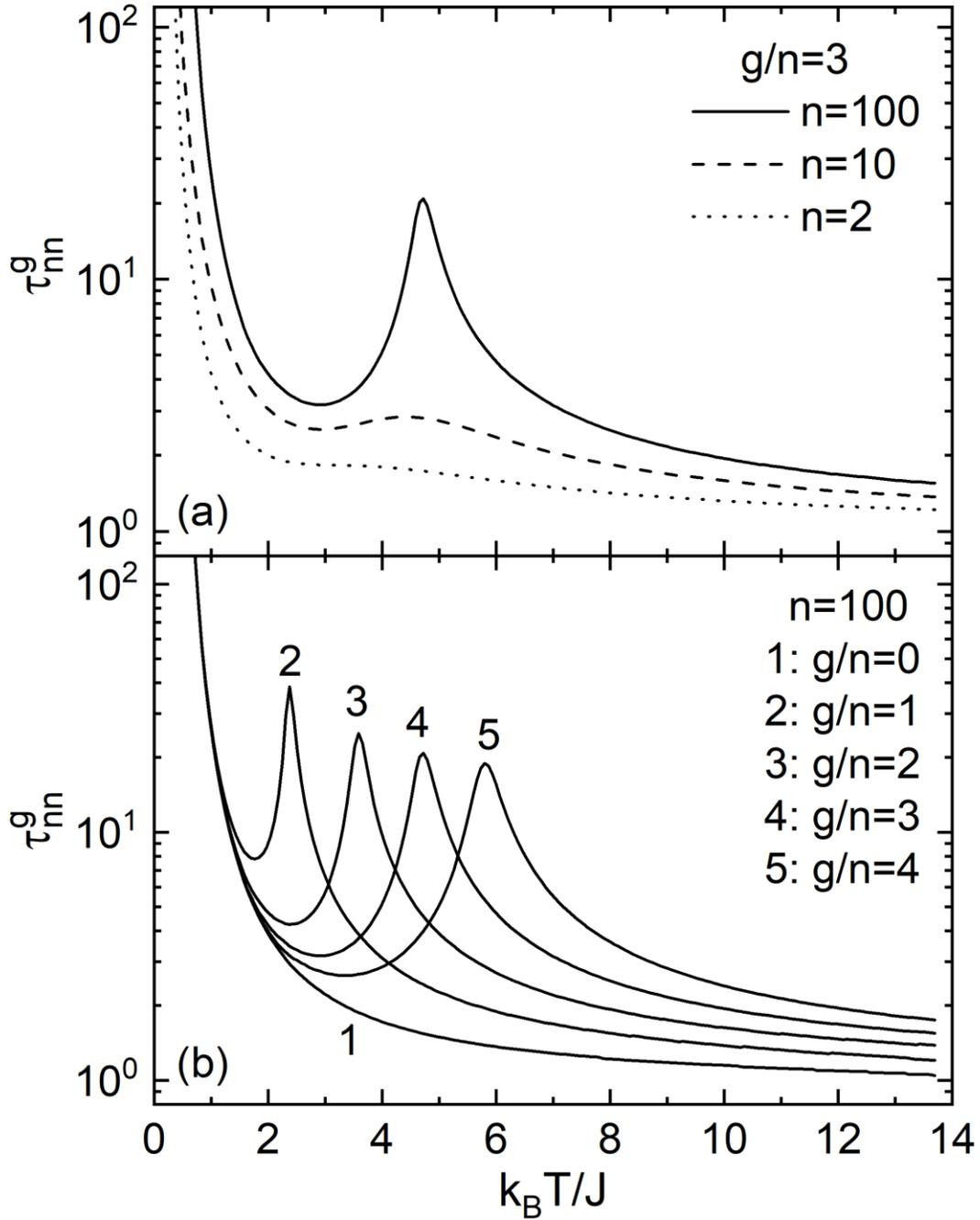

Fig.12 $\nu \tau_{nk}^{g}$ ($k = 1, \cdots n$) of n-g-PSSs vs $T$ for $n = 2, 10, 100$ and $\frac{g}{n} = 3$ (a), and for $n = 100$ and series of $g$ (b).

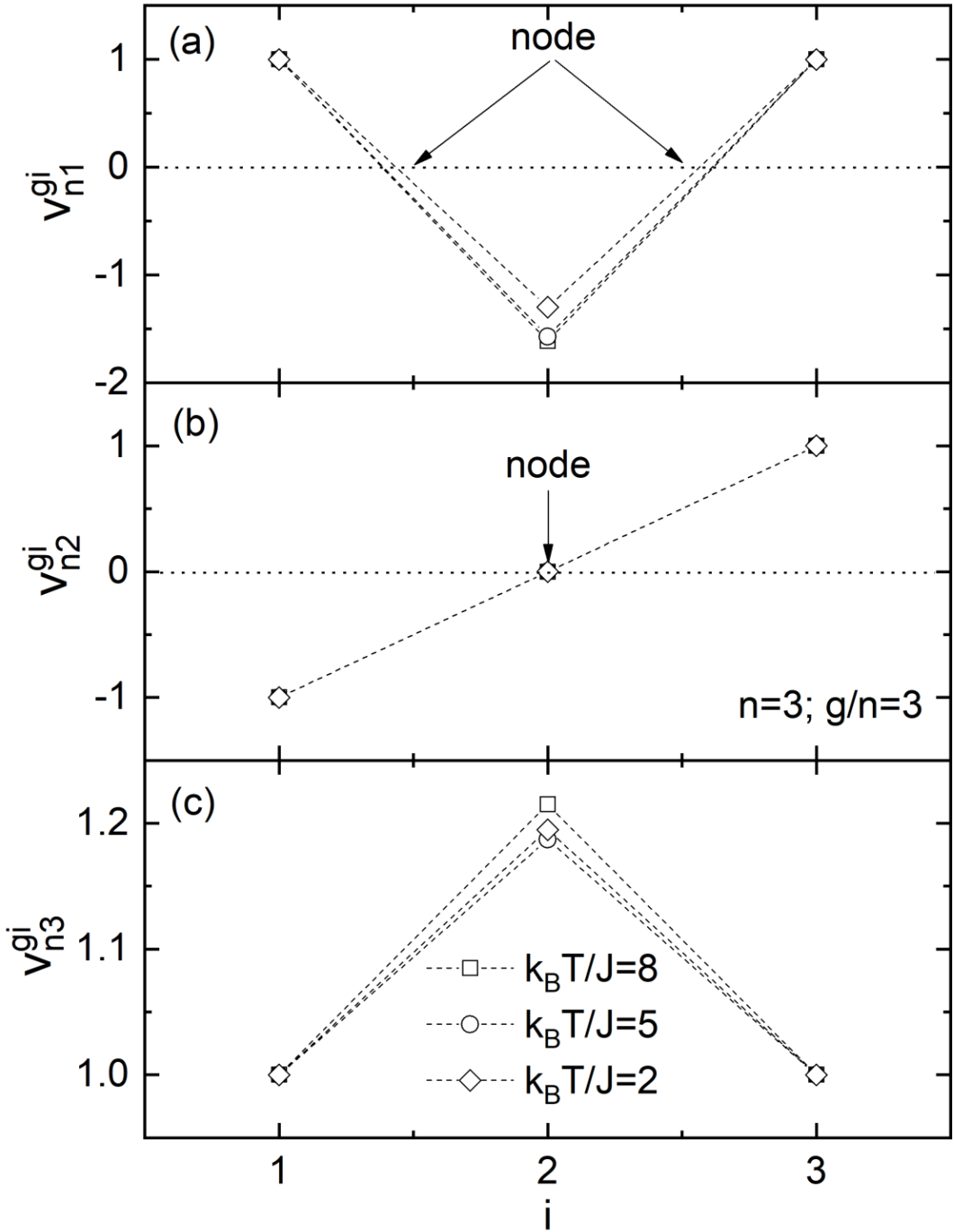

Fig.13 $v_{nk}^{gi}$ ($k = 1, \cdots n$) of SRMs vs PS position ($i = 1, \cdots n$) in n-g-PSSs of $n = 3$.

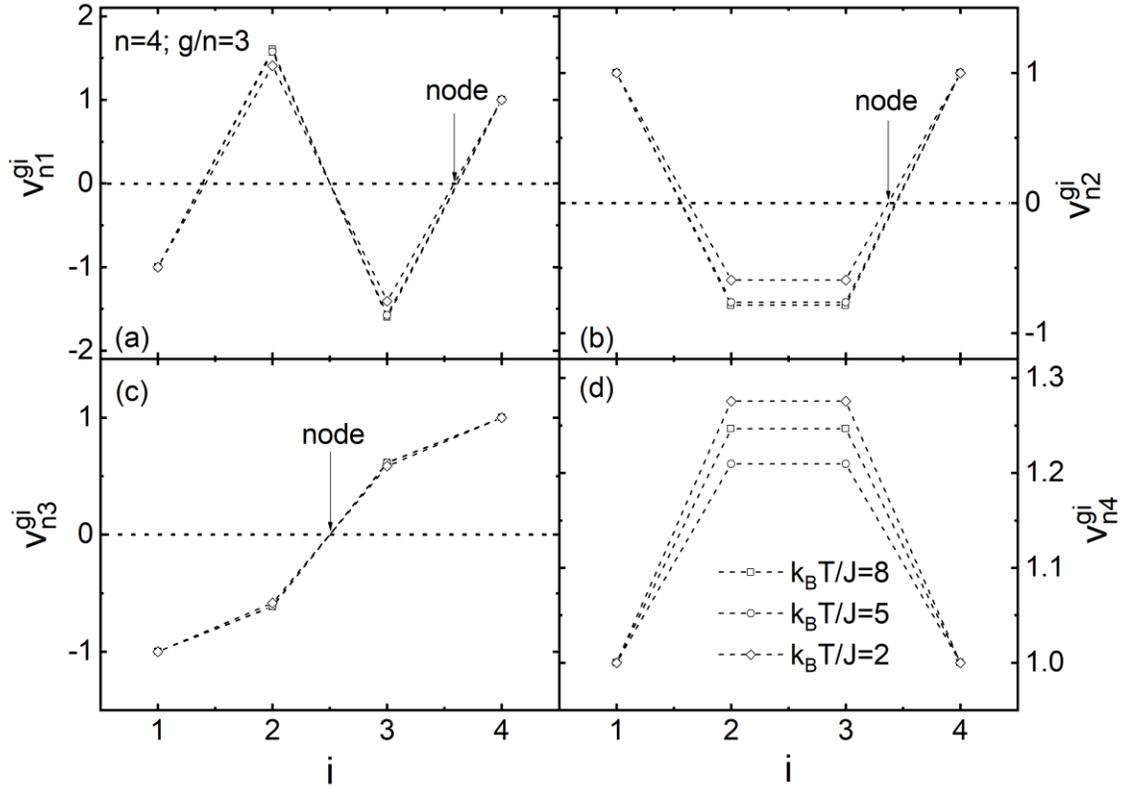

Fig.14 $v_{nk}^{gi}$ $(k = 1, \cdots n)$ of SRMs vs PS position $(i = 1, \cdots n)$ in n-g-PSSs of $n = 4$.

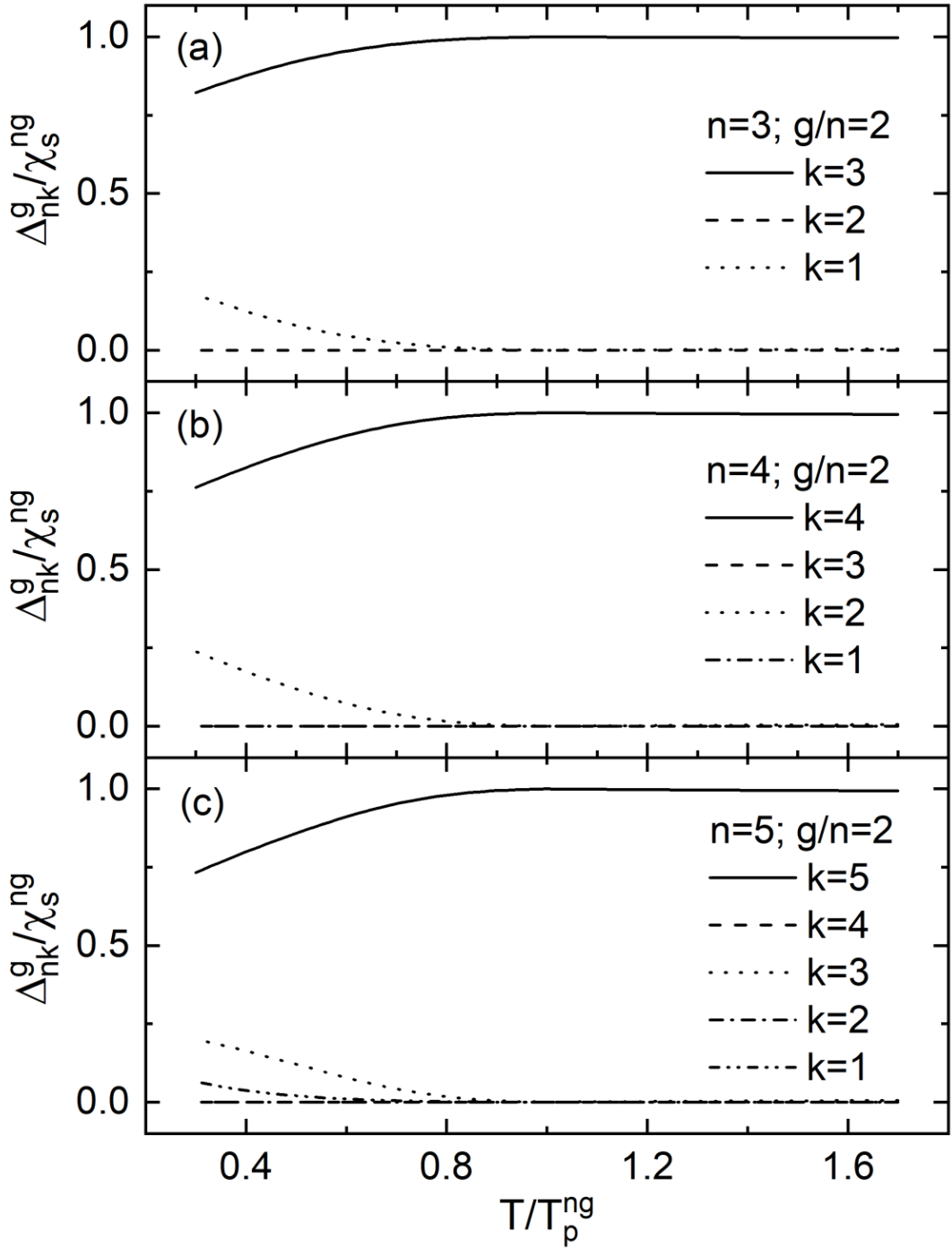

Fig.15 $\Delta_{nk}^{g}/\chi_{s}^{ng}$ $(k=1,\cdots n-1)$ vs $T$ for $n=3,4,5$ and $\frac{g}{n}=2$.

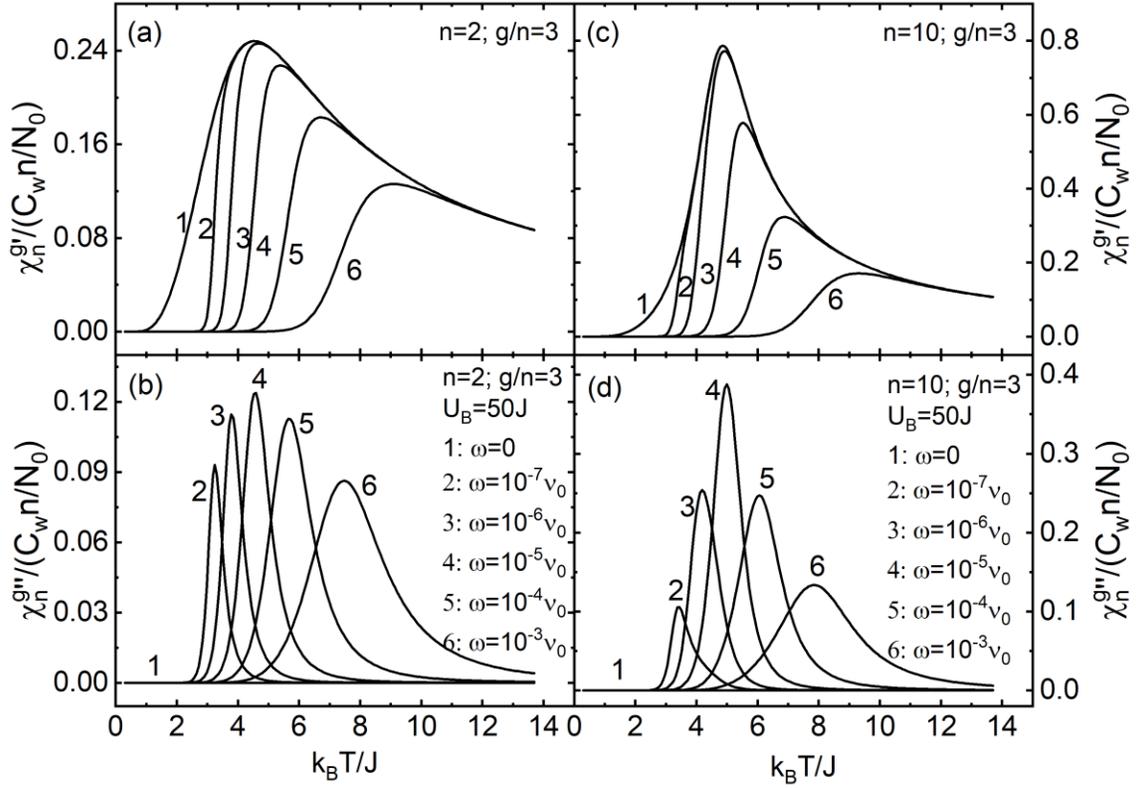

Fig.16 $\chi_n^{g\prime}$ and $\chi_n^{g\prime\prime}$ of n-g-PSSs vs $T$ and $\omega$ when $\frac{g}{n} = 3$, $U_B = 50J$, and $n = 2, 10$.

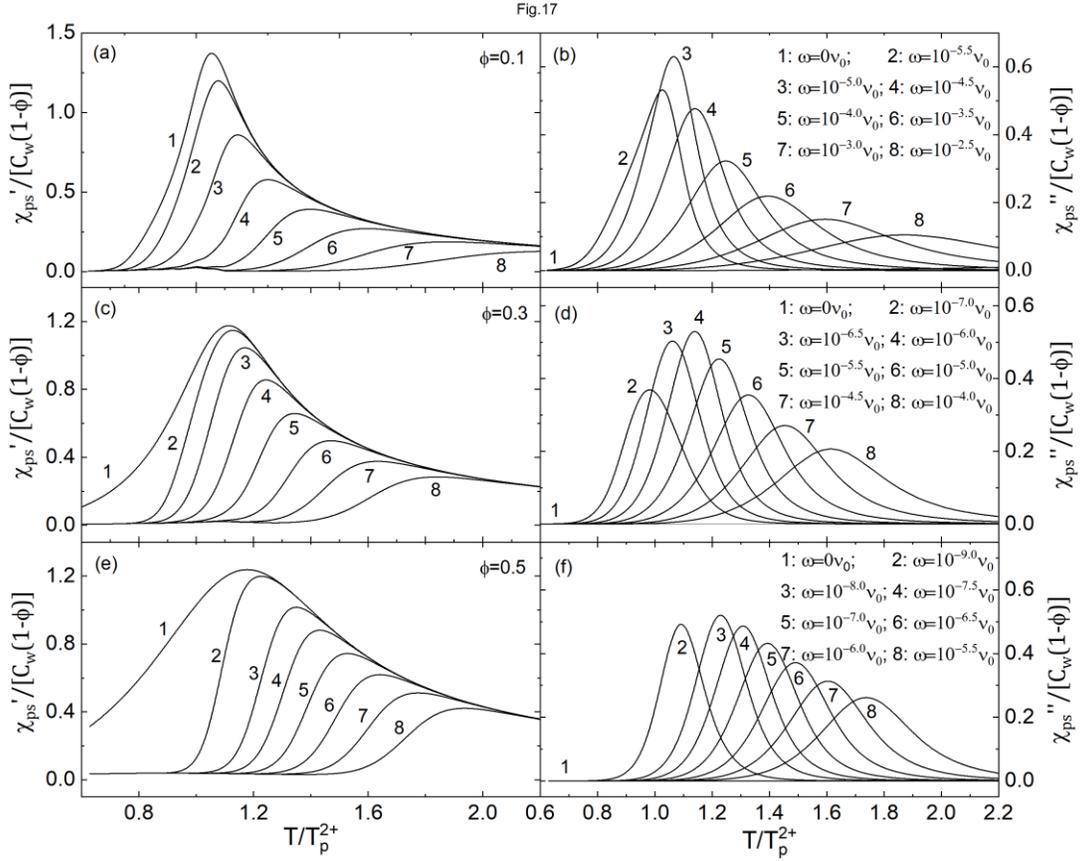

Fig.17 $\chi'_{ps}$ and $\chi''_{ps}$ of 3D-RSIGM vs $T$ and $\omega$ when $U_B = 50J$ and series of $\phi$.

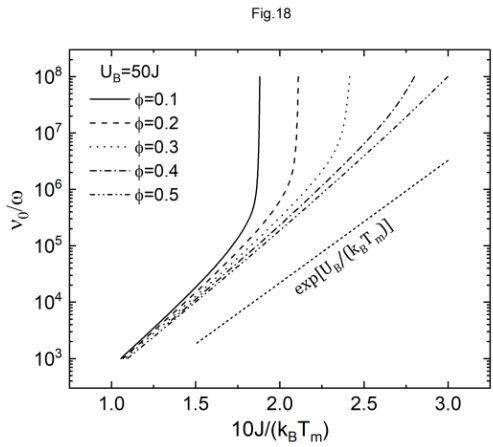

Fig.18 Arrehnius plot of $\omega$ vs $\frac{1}{T_m}$ for 3D-RSIGM when $U_B = 50J$ and series of $\phi$.

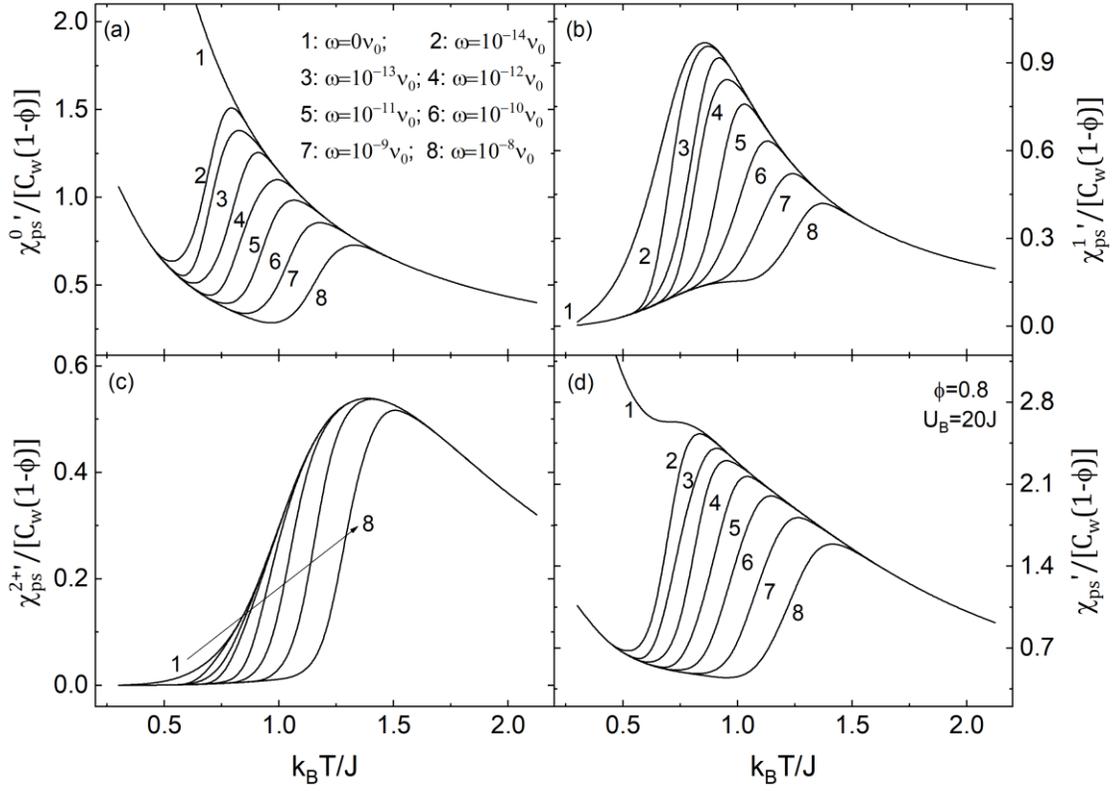

Fig.19 $\chi_{ps}^{0\prime}$, $\chi_{ps}^{1\prime}$, $\chi_{ps}^{2+\prime}$, and $\chi_{ps}^{\prime}$ of 3D-RSIGM vs $T$ and $\omega$ when $U_B = 20J$ and $\phi = 0.7$.

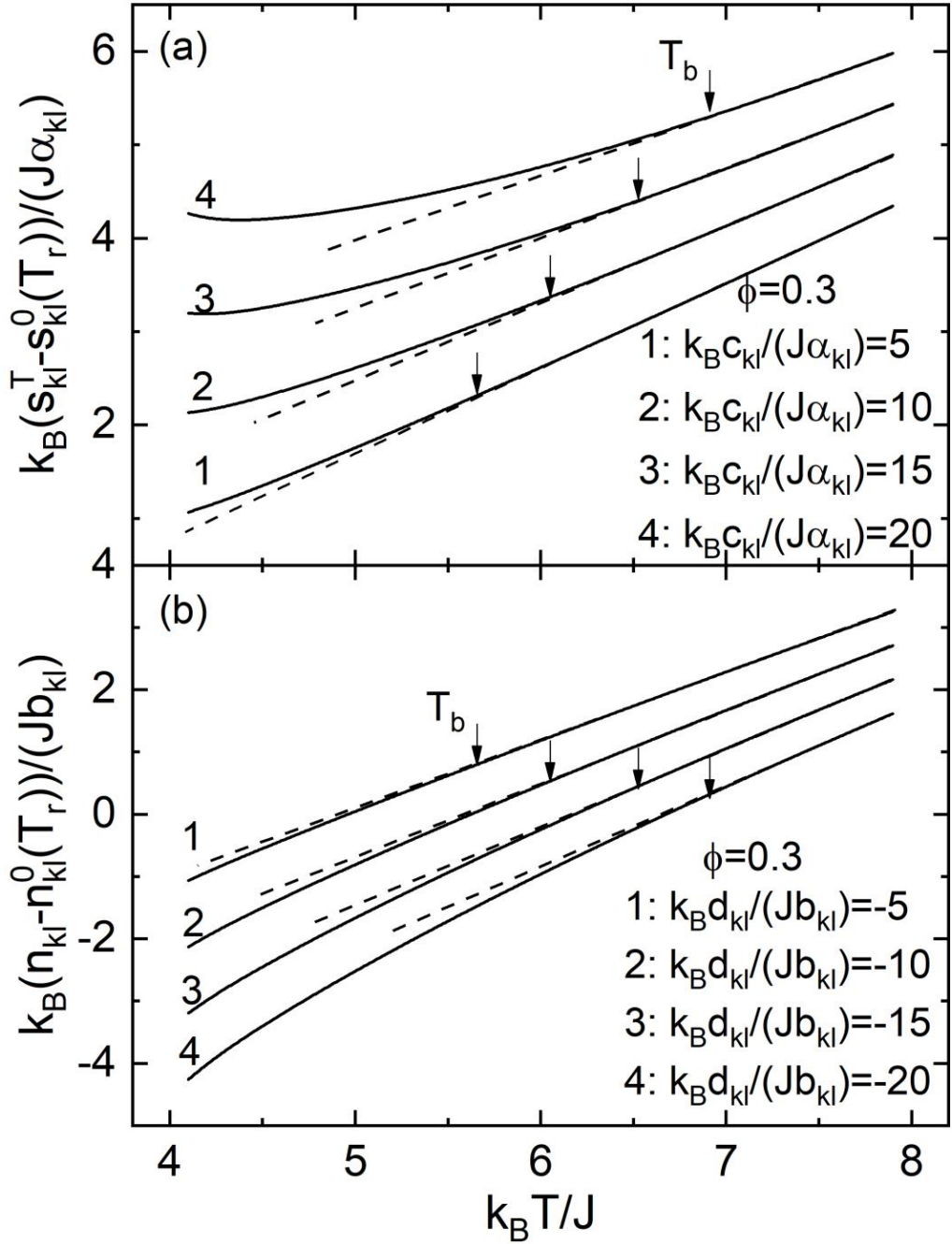

Fig.20 Reduced high temperature $s_{kl}^T$ (a) and $n_{kl}$ (b) of 3D-RSIM vs $T$ for series of $c_{kl}/\alpha_{kl}$ and $d_{kl}/b_{kl}$ respectively.